\documentclass[preprint,authoryear]{elsarticle}

\usepackage[utf8]{inputenc}
\usepackage{amssymb}
\usepackage{xcolor}
\usepackage{tikz}
\usepackage{caption}
\usepackage{subcaption}
\usepackage{amsmath}
\usepackage{url}
\usepackage{makecell}

\definecolor{rwth}   {RGB}{  0  84 159}
\definecolor{rwth-75}{RGB}{ 64 127 183}
\definecolor{rwth-50}{RGB}{142 186 229}
\definecolor{rwth-25}{RGB}{199 221 242}
\definecolor{rwth-10}{RGB}{232 241 250}

\definecolor{black}   {RGB}{  0   0   0}
\definecolor{black-75}{RGB}{100 101 103}
\definecolor{black-50}{RGB}{156 158 159}
\definecolor{black-25}{RGB}{207 209 210}
\definecolor{black-10}{RGB}{236 237 237}

\definecolor{magenta}   {RGB}{227   0 102}
\definecolor{magenta-75}{RGB}{233  96 136}
\definecolor{magenta-50}{RGB}{241 158 177}
\definecolor{magenta-25}{RGB}{249 210 218}
\definecolor{magenta-10}{RGB}{253 238 240}

\definecolor{yellow}   {RGB}{255 237   0}
\definecolor{yellow-75}{RGB}{255 240  85}
\definecolor{yellow-50}{RGB}{255 245 155}
\definecolor{yellow-25}{RGB}{255 250 209}
\definecolor{yellow-10}{RGB}{255 253 238}

\definecolor{petrol}   {RGB}{  0  97 101}
\definecolor{petrol-75}{RGB}{ 45 127 131}
\definecolor{petrol-50}{RGB}{125 164 167}
\definecolor{petrol-25}{RGB}{191 208 209}
\definecolor{petrol-10}{RGB}{230 236 236}

\definecolor{turkis}   {RGB}{  0 152 161}
\definecolor{turkis-75}{RGB}{  0 177 183}
\definecolor{turkis-50}{RGB}{137 204 207}
\definecolor{turkis-25}{RGB}{202 231 231}
\definecolor{turkis-10}{RGB}{235 246 246}

\definecolor{grun}   {RGB}{ 87 171  39}
\definecolor{grun-75}{RGB}{141 192  96}
\definecolor{grun-50}{RGB}{184 214 152}
\definecolor{grun-25}{RGB}{221 235 206}
\definecolor{grun-10}{RGB}{242 247 236}

\definecolor{maigrun}   {RGB}{189 205   0}
\definecolor{maigrun-75}{RGB}{208 217  92}
\definecolor{maigrun-50}{RGB}{224 230 154}
\definecolor{maigrun-25}{RGB}{240 243 208}
\definecolor{maigrun-10}{RGB}{249 250 237}

\definecolor{orange}   {RGB}{246 168   0}
\definecolor{orange-75}{RGB}{250 190  80}
\definecolor{orange-50}{RGB}{253 212 143}
\definecolor{orange-25}{RGB}{254 234 201}
\definecolor{orange-10}{RGB}{255 247 234}

\definecolor{rot}   {RGB}{204   7  30}
\definecolor{rot-75}{RGB}{216  92  65}
\definecolor{rot-50}{RGB}{230 150 121}
\definecolor{rot-25}{RGB}{243 205 187}
\definecolor{rot-10}{RGB}{250 235 227}

\definecolor{bordeaux}   {RGB}{161  16  53}
\definecolor{bordeaux-75}{RGB}{182  82  86}
\definecolor{bordeaux-50}{RGB}{205 139 135}
\definecolor{bordeaux-25}{RGB}{229 197 192}
\definecolor{bordeaux-10}{RGB}{245 232 229}

\definecolor{violett}   {RGB}{ 97  33  88}
\definecolor{violett-75}{RGB}{131  78 117}
\definecolor{violett-50}{RGB}{168 133 158}
\definecolor{violett-25}{RGB}{210 192 205}
\definecolor{violett-10}{RGB}{237 229 234}

\definecolor{lila}   {RGB}{122 111 172}
\definecolor{lila-75}{RGB}{155 145 193}
\definecolor{lila-50}{RGB}{188 181 215}
\definecolor{lila-25}{RGB}{222 218 235}
\definecolor{lila-10}{RGB}{242 240 247}

\usetikzlibrary{arrows.meta}

\usepackage{natbib}
\usepackage{graphicx}
\setcitestyle{authoryear,open={(},close={)}}

\begin{document}
\begin{frontmatter}

\title{To build or not to build - A queueing-based approach to timetable independent railway junction infrastructure dimensioning}
\author[inst1]{Tamme Emunds}
\ead{emunds@via.rwth-aachen.de}

\affiliation[inst1]{organization={Institute of Transport Science,\\ RWTH Aachen},
            addressline={{Mies-van-der-Rohe Straße 1}}, 
            city={Aachen},
            postcode={52074},
            country={Germany}}

\author[inst1]{Nils Nießen}
\begin{abstract}
Many infrastructure managers have the goal to increase the capacity of their railway infrastructure due to an increasing demand. While methods for performance calculations of railway line infrastructure are already well established, the determination of railway junction capacity remains a challenge. This work utilizes the concept of queueing theory to develop a method for the capacity calculation of railway junctions, solely depending on their infrastructure layout along with arrival and service rates. The implementation of the introduced approach is based on advanced model-checking techniques. It can be used to decide which infrastructure layout to build, i.e. whether an overpass for the analysed railway junction is needed. The developed method hence addresses the need for fast and reliable timetable independent junction evaluation in the long-term railway capacity calculation landscape.
\end{abstract}
\end{frontmatter}

\section{Introduction}
The amount of goods and passengers transported by rail is predicted to increase significantly, hence infrastructure managers not only need to build new railway infrastructure but also find possibilities to increase traffic volume on already existing relations. Both of these tasks require sufficient methods to determine the capacity of all sub-units, i.e. lines, junctions and stations, in a railway network.

While Capacity Determination techniques for railway lines, junctions and stations have already been developed and continue to contribute to more efficient rail transportation today, only some methods actually describe timetable independent approaches, allowing for sophisticated infrastructure-dimensioning in early stages of the planning process which is traditionally done in multiple steps \citep{UIC.2013}. Strategical design decisions on a network level may take place decades before operation and are an ongoing process for most infrastructure managers. The technical infrastructure planning does take some time and is done several years in advance, while the definitive timetable construction may be done several months or years prior to operation. Therefore, timetable independent methods determining the capacity of railway infrastructure are of particular importance for the early process-stages including `strategical network design' and `technical infrastructure planning'.

In this work, classical queueing-based capacity analysis methodology is extended to railway junctions by introducing a Continuous-Time Markov Chain formulation based on routes through the infrastructure of a railway junction.
Dependent only on resource conflicts and arrival/service rates of the trains travelling along considered routes, the number of trains waiting for the assignment of their requested infrastructure are estimated by calculating state-probabilities in the introduced Continuous-Time Markov Chain.
To the best of our knowledge, this work is the first to utilize state-of-the-art probabilistic model-checking software to perform the computations on models describing railway junction infrastructure.
The presented approach does not require timetable data to determine the capacity of a railway junction, enabling early-stage infrastructure dimensioning.
Consequently, technical planning decisions, i.e. whether a railway overpass is to be built, can be assisted by capacity calculations with the described approach.
With its versatile range of parameters, multiple applications of capacity research, i.e. infrastructure dimensioning for a fixed traffic demand or capacity determination of a fixed junction infrastructure, may be realised.
In comparison with other approaches to determine timetable independent capacity, the presented method does not rely on simulations or sampling of possible sequences for timetable-compression, but rather on queueing-systems, formulated for multiple service channels.
Further separating this approach from other analytical methods, no deep railway-specific knowledge for a preceding analysis of the considered junction infrastructure is needed, only conflicting routes and either necessary service times or planned operating programs are required as input.

While Section \ref{sec_related_work} includes a literature review regarding other approaches to the determination of railway capacity, Section \ref{sec_methods} introduces formal problem formulations and proposes the new capacity calculation method.
In Section \ref{sec:validation}, the proposed method is compared to a simulation approach with respect to computation times and accuracy of the obtained solutions.
A Case Study, highlighting the applicability of the introduced approach, is performed in Section \ref{sec:Computational_Study}. This work is concluded with a summary and discussion in Section \ref{sec:Discussion}.


\section{Related Work}
\label{sec_related_work}
This section provides an overview regarding the state-of-the-art for railway capacity analysis. After the definition of some key types of performance analysis and frequently used terms, selected examples of literature are matched regarding their associated methodology and capacity definition.

\subsection{Terminology of railway performance analysis}

With their various requirements to the grade of detail, different stages of the planning process can require an analysis of diverse definitions of railway capacity.
While a first definition of railway capacity as the \textit{maximal number of trains traversing a given infrastructure in a given time period under some fixed assumptions} summarizes the concept in a straightforward manner, the interpretation of the included assumptions determines different levels of capacity.
In detail, three types of railway capacity may be distinguished (see also \citealp{Jensen.2020}):
\begin{itemize}
    \item \textit{Theoretical capacity}: The maximum number of requests (i.e. trains, train-route enquiries), that can be scheduled without conflicts on the given infrastructure under consideration of driving dynamics and installed railway control systems.
    \item \textit{Timetable capacity} (sometimes referred to as \textit{maximal capacity}): The maximum number of requests, that can traverse the given infrastructure in acceptable quality when compared to a specified threshold, not only considering driving dynamics, railway control systems, but also operating program specific settings, such as train-mix and arrival processes.
    \item \textit{Operational capacity} (sometimes referred to as \textit{practical capacity}): The maximum number of trains, that can traverse the given infrastructure in acceptable operational quality when compared to a specified threshold, considering driving dynamics, railway control systems, operating programs, and additionally respecting disturbances and (knock-on) delays.
\end{itemize}

Additionally, research has not only been focused on determining the capacity (of any kind), but also on the calculation of the \textit{capacity utilization}
, i.e, the amount of available capacity, consumed in a given timetable.

While methods determining capacity utilization are mostly dependent on a timetable (\textit{timetable dependent}) or at least a fixed order of a given train set, some methods are capable of determining performance indicators without the need for a previously set timetable (\textit{timetable independent}), making them valuable for infrastructure dimensioning in early stages of infrastructure planning processes. Some approaches build on a given timetable and find maximal subsets of a set of trains that can additionally be scheduled (\textit{timetable saturation}), building a category in between both dependency expressions.
In contrast to a timetable, an \textit{operating program} specifies the demand of train types on given lines or routes, hence being indispensable for most performance analyses. 

Depending on the stage of railway infrastructure planning, a timetable may already be available, such that methodologies like \textit{timetable compression}, i.e. routing trains through the infrastructure with a separation of only the minimum headway possible, are useable.

The various railway infrastructure planning stages make use of varying capacity definitions and hence distinct methodologies for the analysis of railway infrastructure. They additionally differ in terms of timetable dependency or granularity of the described infrastructure.
During the following subsection, the methodologies \textit{max-plus-algebra}, \textit{optimisation}, \textit{operational data analysis}, \textit{simulation} and \textit{analytical} are differentiated.

Some additional distinctions of relevant railway performance analysis are made regarding their analysed infrastructure -- lines, junctions, stations or networks --,  their infrastructure decomposition and their utilized solution methods, f.e. \textit{mixed integer programming} (MIP) or matrix calculations.

\subsection{Literature Review}

The described terminology is used in Table \ref{table:lit} to partition some relevant related research regarding their associated category. Additionally, selected utilization, optimisation and delay-propagation methods are briefly introduced in the following, while our contribution is classified along the strongly related analytical approaches. An additional table, categorizing the considered literature can be found in the Appendix in Table \ref{table:lit_all}.

\begin{table}[hp]
\centering
\caption{Literature considering railway performance estimations}
\label{table:lit}
\resizebox{\textwidth}{!}{
\begin{tabular}{c|c|c|c|c|c|c|c}
\thead{methodology} & \thead{capacity \\ type} & \thead{timetable \\ independent}& \thead{microscopic \\ junction \\ evaluations}& \thead {stochastic \\ arrival \\ process} & \thead {stochastic \\ service \\ process} & \thead{application \\ specific \\ infrastructure \\ decomposition} & \thead{solution \\ technique}    \\ \hline \hline
\makecell{timetable \\ compression}           & util                     & no                    & (yes) & no    & no    & no  & compression                              \\ \hline
\makecell{randomized timetable \\ compression}            & util                     & yes                   & (yes) & no    & no    & no  & \makecell{combinatorial \\ optimisation} \\ \hline
max-plus-algebra                              & util                     & (yes)                 & yes   & no    & no    & no  & \makecell{matrix \\ calculations}        \\ \hline
\makecell{timetable \\ saturation}            & theo                     & (yes)                 & (yes) & no    & no    & no  & mostly MIP                               \\ \hline
\makecell{capacity \\ optimisation}           & theo, util               & (yes)                 & (yes) & no    & no    & no  & mostly MIP                               \\ \hline
\makecell{operational \\ data analysis}       & op                       & no                    & no    & (yes) & (yes) & no  & \makecell{stat. analysis \\ and machine learning}  \\ \hline
\makecell{delay propagation \\ analysis}               & op                       & (yes)                 & no    & (yes) & yes   & no  & \makecell{matrix calculations \\ and iterative formula}   \\ \hline
\makecell{simulation}                                    & op                       & no                    & yes   & (yes) & yes   & no  & \makecell{simulation \\ computations}            \\  \hline
\makecell{analytical \\ line capacity}        & tt, op                   & yes                   & no    & yes   & yes   & no  & \makecell{closed formula}  \\ \hline
\makecell{(prior) analytical \\ node capacity}  & tt, op                   & yes                   & yes   & yes   & yes   & yes & \makecell{matrix calculations \\ and iterative formula}    \\ \hline \hline
\makecell{introduced \\ here}                 & tt                       & yes                   & yes   & yes   & yes   & no  & \makecell{model checking}         \\ \hline  \hline
\multicolumn{8}{p{1.7\textwidth}}{Remarks: Please note that a statement in bracelets means that this feature is partially supported, i.e. for some methods within the group or in an incomplete interpretation only. Furthermore, capacity types utilization (util), theoretical (theo), operational (op) and timetable (tt) are abbriviated.}
\end{tabular}
}
\end{table}

Describing the determination of \textbf{capacity utilization} and introducing threshold values for sufficient operational quality, the UIC Code 406 \citep{UIC.2004, UIC.2013} is widely used internationally for capacity assessments on railway lines \citep{Abril.2008, Landex.2009, Goverde.2013} and stations \citep{Landex.2013, Weik.2020}. 
Utilizing the theory of capacity occupation, Max-Plus Automata \citep{Goverde.2007, Besinovic.2018} give measures for the assessment of railway infrastructure and timetable structure.
While traditional compression methods \citep{UIC.2004, Goverde.2007, Abril.2008, Landex.2009, Goverde.2013} require a timetable to calculate capacity consumption, other approaches have been proposed utilizing randomly generated sequences of train types to overcome timetable dependencies 
\citep{Jensen.2017, Jensen.2020, Weik.2020}, focusing on the determination of timetable capacity. An overview and comparison with other capacity consumption methods can be found in \citet{Zhong.2023}.

Furthermore, \textbf{optimisation methods} for the estimation of theoretical capacity have been developed. They mostly formulate (linear) mixed integer programming problems, including approaches for railway lines \citep{Harrod.2009, Yaghini.2014}, stations \citep{Zwaneveld.1996, Zwaneveld.2001, Delorme.2001} and networks \citep{Burdett.2006, Burdett.2015}.
Some approaches rely on solutions to the railway timetabling problem \citep{Cacchiani.2012, Leutwiler.2022}, building timetables utilizing a given infrastructure in a 'best', as defined by objective functions, manner.
Methods may also estimate the capacity occupation of its solution while creating a timetable \citep{Zhang.2016}.
Other approaches are based on the saturation of given timetables, which may also include empty schedules, and optimise the amount of additional traffic \citep{Burdett.2006, Harrod.2009, Liao.2021}.

Going even further, optimisation methods may incorporate rolling stock information \citep{Liao.2021} to the construction of a saturated timetable or estimate the effects of emerging first-order delays to following trains \citep{Mussone.2013}, handling the propagation of so called \textit{knock on} delays.
For this, \citet{Mussone.2013} extend an approach by \citet{Kort.2003}, utilizing max-plus algebras to calculate capacity on lines with some single-track sections, i.e. tunnels.

More detailed insights into operational parameters of specific timetables and detailed infrastructure, rolling stock and delay distribution data can be obtained by utilizing \textbf{simulations}. \citet{Dacierno.2019} provide a comprehensive literature review. While being subject to large computational times, simulations are versatile in their use case. As such, the influence on model performance of different parameters can be analysed, i.e. different buffer time distributions to operational capacity estimations \citep{Zieger.2018}, analysing the propagation of delays on railway infrastructure.

An investigation of \textbf{delay propagation} properties has also been part of further analytical 
\citep{Goverde.2010, Buker.2011} research and subject of machine-learning models \citep{Sahin.2017, Corman.2018}, trained on historical operational data. \citet{Sahin.2017} calculates probabilities for different delay states with the help of a Markov Chain, while \citet{Corman.2018} utilize Bayesian networks. Interested readers are referred to \citet{Spanninger.2023} for a detailed review.

Recent approaches make additional use of \textbf{operational data} to identify capacity bottlenecks. While some research introduces train traffic data mining approaches to analyse actual operation on railway lines \citep{Graffagnino.2012}  or stations \citep{Armstrong.2017}, others \citep{Weik.2022, Corman.2022} discuss the applicability of macroscopic fundamental diagrams. For this, \citet{Weik.2022} introduces a simulation approach to highlight their benefit for further macroscopic research and \citet{Corman.2022} provide an overview of open research applications.



Taking randomly distributed inter-arrival and service times into consideration, \textbf{analytical methods}, based on queueing theory, have been developed for an efficient timetable or operational capacity analysis in the early planning stages.
In \citep{GerhartPotthoff.1970} railway lines and stations are dimensioned, analysing loss probabilities depending on filling and service functions utilizing probability distributions for the arrival process of trains to the analysed station.
While \citet{Schwanhauer.1974, Wakob.1984, Wendler.2007} introduce measures for an analytical determination of the capacity of lines, \citet{Schwanhauer.1978, Niessen.2008, Niessen.2013, Schmitz.2017} analyse junctions, incorporating randomly distributed inter-arrival and service durations.

\citet{Schwanhauer.1974} formulates the STRELE method to calculate expected values for the waiting times of trains in a line infrastructure, which is extended to junction infrastructure in \citet{Niessen.2008}. They hence give measures to calculate the \textbf{operational capacity} of lines and junctions by comparing the estimated waiting times with threshold values \citep{Schwanhauer.1982}.

\citet{GerhartPotthoff.1970, Schwanhauer.1978, Wendler.2007, Wakob.1984, Niessen.2008, Niessen.2013, Schmitz.2017, Weik.2020PhD}, however, implement results for the \textbf{timetable capacity} of the analysed infrastructure by matching corresponding limits with estimated waiting times for trains without taking delays into account.

Calculating the expected waiting times via probabilities for the loss of an arriving train, \citet{GerhartPotthoff.1970, Niessen.2008, Niessen.2013} formulate methods for station \citep{GerhartPotthoff.1970} and junction  infrastructure.
For junction capacity estimations, \citet{Niessen.2008, Niessen.2013} tackles queueing systems with multiple channels, while \citet{Schwanhauer.1978} approximates the waiting times in a junction via a single-channel system.
This single-channel system approximation is based on route-exclusion probabilities and is adapted in \citet{Weik.2020PhD}, joining it with line-capacity advancements in \citet{Wendler.2007}, therefore utilising multi-state service processes and hence more flexible probability distributions when modelling the service process.

\textit{Parameter estimations} have been done \citep{Wakob.1984} to obtain a closed formula for approximating the actual waiting times on a railway line with general independent service and inter-arrival time distributions.




Also directly modelling general independent service and arrival processes, \citet{Schmitz.2017} formulate multi-dimensional Markov Chains and include phase-type distributions in their approximation.
However, their approach is limited to already partitioned junction infrastructure, hence dependent on additional input and deep system knowledge of the algorithm operator. Additionally, they distinguish between train and request types in a queue, resulting in issues with computational memory and scaling, when analysing a more complex infrastructure and/or operating program.

Overall, the methodology landscape is still lacking research combining major advantages of analytical junction capacity methods: Calculating timetable capacity for multi-channel railway junctions while being easily applicable to a broad range of problem formulations, f.e. infrastructure dimensioning for a fixed traffic demand, and utilizing multi-dimensional Markov Chains, enabling the use of advanced queueing theory methodology, such as probabilistic model-checking.

In this work, infrastructure performance is measured by such an approach, analysing timetable capacity by modelling railway junction infrastructure with a multi-dimensional Continuous-Time Markov Chain. This later discussed model features a route-based infrastructure decomposition, hence being easily applicable to more complex junctions, while respecting arising resource conflicts without the need for additional infrastructure partitioning.

\section{Methods}
\label{sec_methods}

\subsection{Junction Layout}
To evaluate the capacity of a railway system, all sub-units of the network need to be assessed. A common differentiation is made between line and node capacity, while methods describing node capacity can additionally be partitioned describing junction and station capacity.

A \textit{railway line} is a connection between two origins, usually equipped with single- or double-track infrastructure. If some traffic shares only part of the route and connects to a third origin, a \textit{railway junction} will be installed to divide the traffic regarding its destination. Unlike at junctions, trains may start and end their journey at \textit{railway stations}. They can be used for passenger- and good-exchange or fulfil primarily operational needs, such as ensuring a possibility for over-takings. 


\begin{figure}[ht]
    \centering
    \includegraphics[width=.9\linewidth]{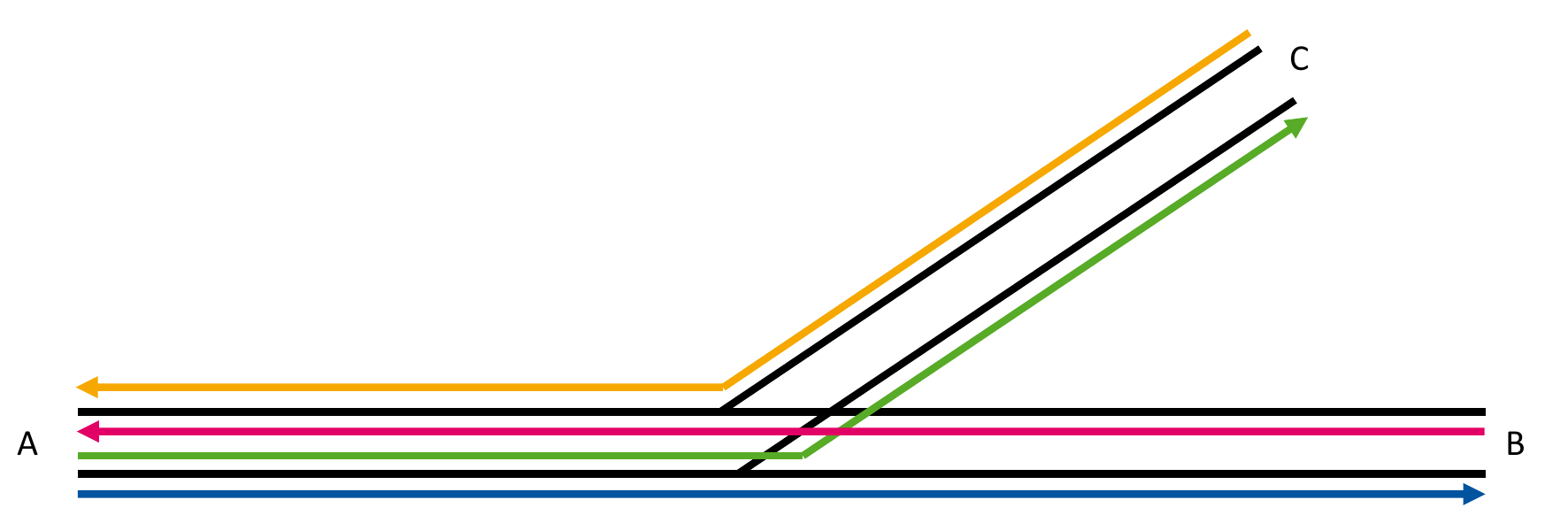}
    \caption{Track-layout for a double-track junction}
    \label{fig:Junctions}
\end{figure}

In Figure \ref{fig:Junctions} an infrastructure designs for a railway junction of two double-track lines is given. The \textit{main line} spans between the two origins A and B, while the
\textit{branching line} connects C with A. 

As in most European countries, we consider operation to be defaulting to right-hand traffic, which suggests the usage of route \textcolor{rwth}{$r_1$} for the direction A to B and routing traffic from B to A via route \textcolor{magenta}{$r_3$} for the main line. Consequently, routes \textcolor{grun}{$r_2$} and \textcolor{orange}{$r_4$} are used accordingly for traffic between A and B.

\begin{table}[h]
\centering
\caption{Routes through the junction}
\label{tab:routes}
\begin{tabular}{l|l|l|l}
Name                        & Origin     & Destination & Conflicts       \\ \hline
\textcolor{rwth}{$r_1$}     & A     & B       & \textcolor{grun}{$r_2$}          \\
\textcolor{grun}{$r_2$}     & A     & C  & \textcolor{rwth}{$r_1$}, \textcolor{magenta}{$r_3$}    \\
\textcolor{magenta}{$r_3$}  & B      & A      & \textcolor{grun}{$r_2$}, \textcolor{orange}{$r_4$}      \\
\textcolor{orange}{$r_4$}   & C & A      & \textcolor{magenta}{$r_3$}
\end{tabular}
\end{table}

When dividing traffic on the junction into routes, those four different trails may be considered for both junction layouts (Table \ref{tab:routes}). The infrastructure can be abstracted along those routes to model its operation.

\subsection{Problem Formulation}
\label{subsection_problem_formulation}

A railway junction infrastructure $J = (R, C)$ consists of a set $R$ of $k$ routes $r \in R$ and a \textit{conflict-matrix} $C \in \{0,1\}^{k \times k}$. Two routes $r_i, r_j \in R$ are described as \textit{conflicting}, i.e. they cannot be used at the same time, by denoting $C_{i,j} = 1$, and as not conflicting by $C_{i,j} = 0$. Per default, the same route cannot be used twice at a time, hence $C_{i,i} = 1$ for all $i \in \{1, \dots, k\}$.

An \textit{operating program} on the railway junction $J$ is a set $A$ of \textit{demands} $a = (r,n)$ corresponding to a route $r \in R$ and a total number of trains $n \in \mathbb{N}$ of that demand in the time horizon $U$. In order to service a train of demand $a=(r,n) \in A$, the service time $t_{\text{service}}(a)$ may be calculated using microscopic railway software. The service time of demand $a$ is dependent on the infrastructure and safety technology in $J$ (and adjacent railway lines and junctions), as well as some train-specific characteristics, such as braking percentages, acceleration capabilities, or total mass.

The problem of dimensioning of railway junctions can be formulated by two inverse statements:

\begin{enumerate}
    \item[(I)] Given a railway junction $J$ and a set of possible operating programs $\{A_1, \dots , A_l\}$. Which operating program $\hat{A} \in \{A_1, \dots , A_l\}$ is the \textit{largest} to be able to be completed on the junction $J$ in acceptable quality? 

    \item[(II)] Given an operating program $A$ and a set of possible infrastructure layouts $\{J_1, \dots J_l\}$. Which infrastructure $\hat{J} \in \{J_1, \dots J_l\}$ is the most affordable, sufficient for acceptable quality in the completion of the desired operation program?
\end{enumerate}

Note that for both problem statements an ordering of some sort can be given for the set of possible solutions. The set of possible operating programs $\{A_1, \dots , A_l\}$ can be sorted by the total number of trains in the operating program for statement (I).
Hence, the \textit{largeness} (as in statement (I)) of an operating program may be evaluated by a given order.
Regarding statement (II), the volume of funds needed for the construction of the infrastructure layouts may be the metric for the set of possible infrastructure layouts $\{J_1, \dots J_l\}$.

While problem statement (I) can be used to assess the theoretical maximal capacity of a railway junction, railway infrastructure operators could be mostly interested in solutions to the statement (II): The operational program can often be assessed beforehand, i.e. when fixed by external factors, such as governments, promoting a modal shift to railways \citep{EuropesRail.2015}, or increasing demands when establishing new industry locations.

With the described approach, both problem statements may be solved by using the infrastructure and operating program in the formulation of a queueing system and analyzing its characteristics.

\subsection{Queueing System}
\label{sec:queueing_system}

Queueing theory \citep{Bolch.2006, Zukerman.2013} has been extensively used to describe the processes in a transportation system.
An analysed entity is usually divided into a \textit{service} and a \textit{waiting} part (\textit{queue}).
Incoming duties may either be assigned to a processing \textit{channel} or, if none is available, to the next slot in the queue.
The Kendall notation \citep{Kendall.1953} has been developed to abbreviate definitions of Queueing Systems with
 \begin{center}
     $A$/$B$/$n$/$m$.
 \end{center}
Arrival ($A$) and service processes ($B$) can be modelled with arbitrary probability distributions. The described Queuing System can contain one or more service channels ($n$) and any natural number of waiting slots ($m$) - including none or infinitely many. 
In this work, Exponential ($M$) and General independent ($GI$) distributions are considered to describe arrival and service processes, but generally, other probability distributions can be utilized as well.

Modelled Queueing Systems are analysed by a set of relevant parameters.
For \textit{Markovian} (or \textit{Exponential distributed}) models (\textit{Markov Chains}), the state at any point in time is the only dependency to the future evolution of the process (see \citet[Ch. 2.4]{Zukerman.2013}).
Hence, transition rates between the states in a Markov Chain can be given. We denote \textit{arrival rates} by $\lambda$ and \textit{service rates} by $\mu$. They can be calculated with the use of expected values for the inter-arrival time $ET_A$ \begin{equation}
    \lambda = \frac{1}{ET_A},
\end{equation}
and service times $ET_S$
\begin{equation}
    \mu = \frac{1}{ET_S}.
\end{equation}
Further, we describe the \textit{occupancy rate} of a queuing system by
\begin{equation}
\label{rho_def}
    \rho = \frac{\lambda}{\mu}.
\end{equation}
Furthermore, variance coefficients of the arrival process $v_A$ and of the service process $v_B$ may be used for estimating supplementary characteristics.
Additional parameters include the \textit{estimated length of the queue} $EL_W$ and the \textit{probability of loss} $p_{loss}$, describing the probability that incoming requests can not be considered in the system, as all service and waiting slots may be occupied.

An example of the use of a Queueing System is the determination of the capacity of a railway line \citep{Wakob.1984, Wendler.2007} , which is usually described with a $GI/GI/1/\infty$ system. As an analytical solution to those systems has not yet been found, one can model it as an $M/M/1/\infty$ system and use an approximation formula \citep{Gudehus.1976} for the calculation of the expected queue length
\begin{equation}
\label{solution_mm1inf}
    EL_W(GI/GI/1/\infty) \approx  EL_W(M/M/1/\infty) \cdot c  =\frac{\rho^2}{1-\rho} \cdot \frac{v_A^2 + v_B^2}{2}
\end{equation}
with a factor $c = \frac{v_A^2 + v_B^2}{2}$.

Further approximations for the lengths of the queue in general independent arrival and/or service processes have been developed in \citet{Wakob.1984, KlausFischer.1990, Wendler.2007, Weik.2020PhD}.

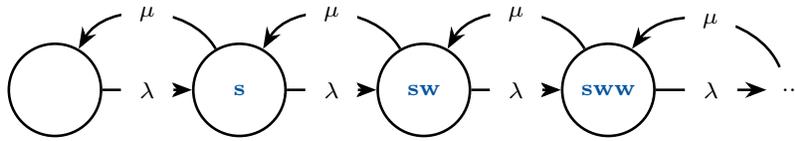
\begin{figure}[ht]
    \centering
    \resizebox{.9\linewidth}{!}{\begin{tikzpicture}
    \begin{scope}[every node/.style={circle,thick,draw, font=\scriptsize},
                    state/.style={minimum size=1cm}]
        \node [state] (s) at (0,0) {};
        \node [state] (b2) at (2,0) {\textbf{\textcolor{rwth}{s}}};
        
        \node [state] (b2w2) at (4,0) {\textbf{\textcolor{rwth}{s}\textcolor{rwth}{w}}} ;
        
        \node [state] (b2w2w2) at (6,0) {\textbf{\textcolor{rwth}{s}\textcolor{rwth}{w}\textcolor{rwth}{w}}} ;
        
        \node [draw=none] (b2w2w2w2) at (8,0) {...} ;
        
    \end{scope}
    \begin{scope}[>={Stealth[black]},
              every node/.style={fill=white,circle},
              every edge/.style={draw=black, thick, font=\scriptsize}]
        \path [->] (s) edge node {$\lambda$} (b2);
        
        \path [->] (b2) edge node {$\lambda$} (b2w2);
        
        \path [->] (b2w2) edge node {$\lambda$} (b2w2w2);

        \path [->] (b2)[bend right=60] edge node {$\mu$} (s) ;
        
        \path [->] (b2w2)[bend right=60] edge node {$\mu$} (b2) ;

        \path [->] (b2w2w2)[bend right=60] edge node {$\mu$} (b2w2) ;
        
        \path [->] (b2w2w2) edge node {$\lambda$} (b2w2w2w2);
        \path [->] (b2w2w2w2)[bend right=60] edge node {$\mu$} (b2w2w2) ;
        
    \end{scope}
\end{tikzpicture}}
    \caption{Markov Chain for a $M/M/1/\infty$ Queueing System}
    \label{fig:Markov_m_infty}
\end{figure}

Figure \ref{fig:Markov_m_infty} introduces a graphical representation of a Continuous-Time Markov Chain (see \citet{Ross.2014} for a definition) modelling the $M/M/1/\infty$ system. It consists of a root state (without a label) as well as states with one currently serviced unit (denoted by 's') and some units in the queue (each denoted by 'w'). Ordering the states by the number of units in the system, as of being serviced and currently waiting for units, transitions between subsequent states correspond to the arrival of a unit into the Queueing System (denoted by the arrival rate '$\lambda$') or the termination of the service of a unit (denoted by the service rate '$\mu$').

The following Section generalizes this concept to railway junctions, which allow more complex resource allocations than on a railway line without multiple possible routes.

\subsection{Modelling railway junctions}
\label{section_modelling}

In this work, railway junctions are considered, which differ from railway lines in one central aspect: It may be feasible for multiple trains to use the infrastructure at the same time - depending on the routes, they are scheduled to take.

\subsubsection{States}
Modelling those infrastructures as a Continuous-Time Markov Chain $MC = (S, T)$, used states $s \in S$ need to be distinguishable regarding the currently serviced and waited-for route(s).

The general set of possible states 
\begin{equation}
    \hat{S} = \left\{\left(q_1, b_1, \dots, q_k, b_k\right) | q_i \in \{0, \dots, m\}, b_i \in \{0,1\} \right\}
\end{equation}
can be obtained by utilizing the information in the set of routes $R$ of a junction $J = (R,C)$. A State $s=\left(q_1, b_1, \dots, q_k, b_k\right) \in \hat{S}$ contains information regarding the number of trains $q_i \in \{0, \dots, m\}$ waiting in the queue for route $r_i$ and whether a route $r_i$ is currently services $b_i \in \{0,1\}$.

The state-space $\hat{S}$ can be further restricted to 
\begin{equation}
    S = \left\{\left(q_1, b_1, \dots, q_k, b_k\right) \in \hat{S} | \sum_{i=1}^k \sum_{j=1}^{k} \left(C_{i,j} b_i b_j\right) = 0 \right\}
\end{equation}
by applying the conflicts described in the conflict-matrix $C$. Furtheron, entries $q_i$ or $b_i$ in a State $s=\left(q_1, b_1, \dots, q_k, b_k\right) \in S$ are also referenced by $q_{s,i}$ or $b_{s,i}$.

\subsubsection{Transitions}
Transitions $(u,v) = t \in T$ between two states $u,v \in S$ can correspond to either the \textit{arrival} of a train for route $r_i$, the \textit{completion of service} on route $r_i$ or the \textit{choice} which holding train to service next.

The used transition rates of arrival-transitions are given by the arrival rate $\lambda_i$, of service-transitions by the service rate $\mu_i$, and of choice-transitions by the maximum rate $M$. Those transition rates can be obtained by the described operational program $A$ and the time horizon $U$.

An arrival transition between
\begin{equation}
    u=\left(q_1, b_1, \dots, q_i, b_i, \dots, q_k, b_k \right)
\end{equation}
and
\begin{equation}
    v=\left(q_1, b_1, \dots, q_i + 1, b_i, \dots, q_k, b_k \right)
\end{equation}
utilizes the arrival rate
\begin{equation}
    \lambda_i = \frac{\sum_{(r_i, n) \in A} (n)}{U},
\end{equation}
where the number of trains using route $r_i$, $\sum_{(r_i, n) \in A} (n)$, is divided by the time horizon $U$ .

Transitions for a service process can be modeled between
\begin{equation}
    u=\left(q_1, b_1, \dots, q_i, 1, \dots, q_k, b_k \right)
\end{equation}
and
\begin{equation}
    v=\left(q_1, b_1, \dots, q_i, 0, \dots, q_k, b_k \right),
\end{equation}
utilizing the service rate
\begin{equation}
    \mu_i = \frac{1}{\frac{\sum_{(r_i, n) \in A} (t_{\text{service}}((r_i,n)) \cdot n)}{\sum_{(r, n) \in A} (n)}},
\end{equation}
which corresponds to the reciprocal of the average service times of all routes, weighted by the amount of trains on each route.

Choice-transitions $t = (u,v)$ exclusively start at states $u \in S$ with
\begin{equation}
    \sum_{i=1}^{k} b_{u,i} = 0
\end{equation}
and at least two conflicting sets of routes $R_i, R_j \subset R$ , with
\begin{equation}
\begin{split}
      q_{u,o} > 0, ~\forall r_o \in R_i \\
    q_{u,p} > 0 , ~\forall r_p \in R_j ,
\end{split}
\end{equation}
which are not operable simultaneously, and end at states $v \in S$, with at least one serviced route. They correspond to the choice of which route to operate next when multiple options are possible.

Since the choice-transitions should not induce additional time in the system, the maximum rate $M$ should be large enough such that the additional expected time in the system per choice-transition $1/M$ is sufficiently small. In this work, choice-transitions with identical transition rates are included to model the different decisions between the route(s) to be serviced next. Hence, the obtained results can be assumed to be independent of disposition strategies. The maximum rate is further chosen as $M=600$, corresponding to a rate of 10 per second, which is deemed a competitive approximation.

\subsubsection{Example}
\label{sec_example}
\begin{figure}[ht]
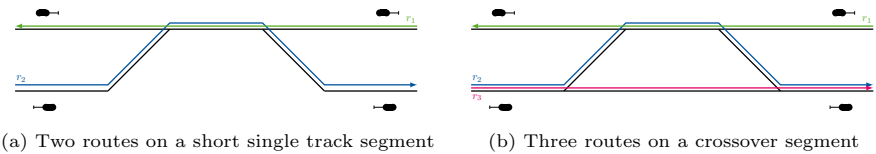

\centering
\begin{subfigure}[t]{.5\textwidth}
  \centering
  \resizebox{.9\linewidth}{!}{\input{infrastructure_Markov_example.tikz}}
  \caption{Two routes on a short single track segment}
  \label{fig:conflicting_routes_single_track}
\end{subfigure}%
\begin{subfigure}[t]{.5\textwidth}
  \centering
  \resizebox{.9\linewidth}{!}{\input{infrastructure_Markov_example2.tikz}}
  \caption{Three routes on a crossover segment}
  \label{fig:conflicting_routes_crossover}
\end{subfigure}
\caption{Conflicting routes in different infrastructure}
\label{fig:conflicting_routes}
\end{figure}

The introduced  Continuous-Time Markov Chain $MC = \left(S, T\right)$ can be used to model Queueing Systems with more complex service regulations. In Figure \ref{fig:conflicting_routes}, two different examples of railway track layouts are presented. While layout \ref{fig:conflicting_routes_single_track} corresponds to a short single-track segment with two possible routes, layout \ref{fig:conflicting_routes_crossover} illustrates the infrastructure of a crossover segment with three routes, two of which (\textcolor{grun}{$r_1$} and \textcolor{magenta}{$r_3$}) being operable simultaneously.

\begin{figure}[ht]
  \centering
  \resizebox{.5\linewidth}{!}{
        

\begin{tikzpicture}
    \begin{scope}[every node/.style={circle,thick,draw,font=\scriptsize},
                    state/.style={minimum size=1cm}]
        \node [state] (s) at (0,0) {};
        \node [state] (b1) at (4,4) {\textbf{\textcolor{grun}{s}}};
        \node [state] (b2) at (4,-4) {\textbf{\textcolor{rwth}{s}}};

        \node [state] (b1w1) at (8,6) {\textbf{\textcolor{grun}{s}\textcolor{grun}{w}}};
        \node [state] (b1w2) at (8,2) {\textbf{\textcolor{grun}{s}\textcolor{rwth}{w}}};
        \node [state] (b2w1) at (8,-2) {\textbf{\textcolor{rwth}{s}\textcolor{grun}{w}}} ;
        \node [state] (b2w2) at (8,-6) {\textbf{\textcolor{rwth}{s}\textcolor{rwth}{w}}} ;

        \node [state] (b1w1w2) at (12,4) {\textbf{\textcolor{grun}{s}\textcolor{grun}{w}\textcolor{rwth}{w}}};
        \node [state] (b2w1w2) at (12,-4) {\textbf{\textcolor{rwth}{s}\textcolor{grun}{w}\textcolor{rwth}{w}}} ;

        \node [state] (w1w2) at (12,0) {\textbf{\textcolor{grun}{w}\textcolor{rwth}{w}}} ;
        
    \end{scope}
    \begin{scope}[>={Stealth[black]},
              every node/.style={fill=white,circle},
              every edge/.style={draw=black, very thick}]
        \path [->] (s) edge node {$\lambda_1$} (b1);
        \path [->] (s) edge node {$\lambda_2$} (b2);
        
        \path [->] (b1) edge node {$\lambda_1$} (b1w1);
        \path [->] (b1) edge node {$\lambda_2$} (b1w2);
        \path [->] (b2) edge node {$\lambda_1$} (b2w1);
        \path [->] (b2) edge node {$\lambda_2$} (b2w2);

        \path [->] (b1)[bend right=60] edge node {$\mu_1$} (s) ;
        \path [->] (b2)[bend left=60] edge node {$\mu_2$} (s) ;

        \path [->] (b1w1)[bend right=60] edge node {$\mu_1$} (b1) ;
        \path [->] (b1w2) edge node {$\mu_1$} (b2) ;
        \path [->] (b2w1) edge node {$\mu_2$} (b1) ;
        \path [->] (b2w2)[bend left=60] edge node {$\mu_2$} (b2) ;

        \path [->] (b1w1) edge node {$\lambda_2$} (b1w1w2);
        \path [->] (b1w2) edge node {$\lambda_1$} (b1w1w2);
        \path [->] (b2w1) edge node {$\lambda_2$} (b2w1w2);
        \path [->] (b2w2) edge node {$\lambda_1$} (b2w1w2);
        
        \path [->] (b1w1w2) edge node {$\mu_1$} (w1w2);
        \path [->] (b2w1w2) edge node {$\mu_2$} (w1w2);
        
        \path [->] (w1w2) edge node {$M$} (b2w1);
        \path [->] (w1w2) edge node {$M$} (b1w2);
        
    \end{scope}
    \matrix [draw, above right] at (current bounding box.south west) {
        \node [label=right:service $r_1$] {\textbf{\textcolor{grun}{s}}};\\
        \node [label=right:service $r_2$] {\textbf{\textcolor{rwth}{s}}};\\
        \node [label=right:waiting for $r_1$] {\textbf{\textcolor{grun}{w}}};\\
        \node [label=right:waiting for $r_2$] {\textbf{\textcolor{rwth}{w}}};\\
    };
\end{tikzpicture}}
  \caption{Markov Chain modelling the Queueing System for the single track segment}
  \label{fig:markov_single_track}
\end{figure}
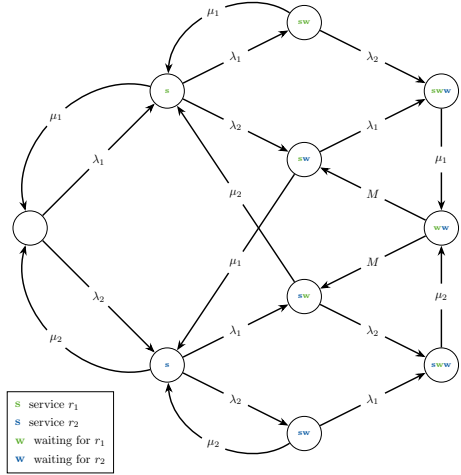%

\begin{figure}[ht]
  \centering
  \resizebox{.9\linewidth}{!}{\input{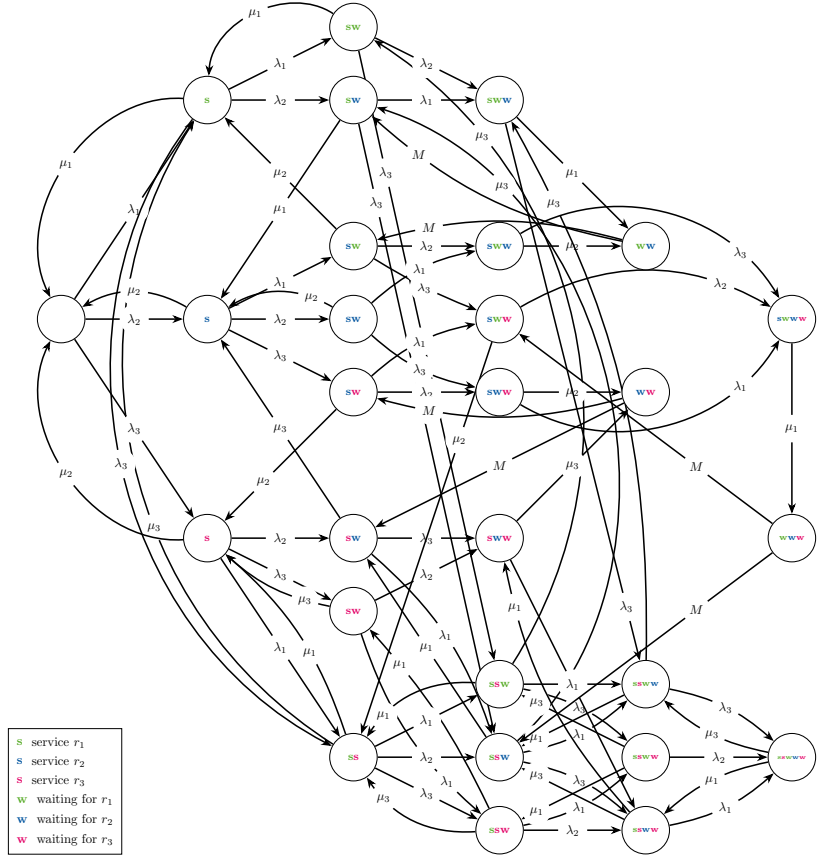}}
  \caption{Markov Chain modelling the Queueing System for the crossover segment.}
  \label{fig:markov_crossover}
\end{figure}

Figures \ref{fig:markov_single_track} and \ref{fig:markov_crossover} give graphical representations of continuous-time Markov Chains modelling the single track segment (Figure \ref{fig:conflicting_routes_single_track}) and the crossover segment (Figure \ref{fig:conflicting_routes_crossover}).
Starting at the most left state, arrivals to their two or three routes $r_i$ are possible with their respective arrival rate $\lambda_i$.
Since the Queueing System is empty, an arriving train starts with its service time immediately after arriving.
While every other train, arriving in the single track system (Figure \ref{fig:conflicting_routes_single_track}) during the service time of the first train, will be allocated to a waiting slot, the assignment of a second train in the Queueing System for the crossover segment does depend on the used routes.
If the first train uses \textcolor{grun}{$r_1$} or \textcolor{magenta}{$r_3$} and the second train the other of both, the second train can be serviced immediately.
If the two trains share the same route or one of them is using route \textcolor{rwth}{$r_3$}, the second train has to be assigned to a waiting slot.

Those implications of route exclusions continue in both graphs and all combinations of arriving and serviced trains are modeled with one limitation: 
While in theory, unlimited state spaces and therefore the modelling of $m=\infty$ waiting positions are possible, an analysis of those models is very complex and in practical operation, the number of waiting trains will always be limited due to space restrictions. Hence, the examples in Figure \ref{fig:markov_single_track} and \ref{fig:markov_crossover} contain only one waiting slot per route. Trains arriving to a route while all waiting slots are occupied are therefore neglected. 

\begin{table}[ht]
\centering
\caption{Number of states by the number of waiting positions in Queuing Systems for the examples in Figure \ref{fig:conflicting_routes}.}
\label{tab:state_numbers}
\begin{tabular}{l|l|l|l|l|l}
number of waiting positions $m$ & 1  & 2   & 4   & 8    & 16    \\ \hline
single track segment        & 10 & 23  & 67  & 227  & 835   \\ \hline
crossover segment           & 28 & 89 & 397 & 2261 & 15013
\end{tabular}
\end{table}

Table \ref{tab:state_numbers} lists the number of states for the two models, rising with the number of waiting positions $m$. Since the probability of loss $p_{loss} = p_{loss}(m)$ is decreasing with the number of waiting positions per route, a sufficient number of waiting positions may be calculated by identifying a limit value $p^{\ast}_{loss}$ and determining the lowest $m \in \mathbb{N}$ satisfying
\begin{equation}
    p_{loss}(m) \leq p^{\ast}_{loss}
\end{equation}
iteratively or by testing relevant expressions of $m$.

A Model for the junction in Figure \ref{fig:Junctions} does incorporate more complex route conflicts and additional parameters, leading to substantial higher state numbers, i.e. $|S| = 128$ for an instance with only one waiting slot $m=1$ per route. Hence, we refrain from including a graphical representation of the induced Continuous-Time Markov Chain. In the online repository \citep{TammeEmunds.2023} however, a full description of the used model, obtained with the definition in Section \ref{section_modelling}, is given.

\subsection{Determination of the estimated length of the queue}

While closed-form analytical solutions exist for the estimated length of the queue $E_{LW}$ for some Queueing Systems (i.e. equation (\ref{solution_mm1inf}), more examples in \citet{KlausFischer.1990}), the analysis of more complex systems remains challenging. Since Continuous-Time Markov Chains may be used to model the behavior of probabilistic programs, the calculation of state-probabilities is a substantial part of the verification of software systems. 
Furthermore, special software tools have been developed to perform the so-called \textit{probabilistic model-checking}, therefore including sophisticated algorithms to calculate state possibilities and enabling higher-order evaluations. This work utilizes the model checker \textit{Storm} \citep{Hensel.2022}
.

Storm parses Markov Chain models, i.e. formulated in the \textit{PRISM modelling language} \citep{ DaveParkerGethinNormanMartaKwiatkowska.2000, KNP11}, and gives tools to check specified \textit{properties} on them. Properties are formulas describing paths on sets of states of a Markov Chain and can be used to calculate probabilities or reachability statements of states. This process is described as probabilistic model-checking and Storm utilizes different \textit{engines} to perform this task, automatically detecting the most suitable engine for the specified input. 

For this work, the introduced Continuous-Time Markov Chain model has been declared in the PRISM modelling language  for multiple infrastructure layouts and read into Storm for the calculations of the expected length of the queue of each route. The formulated model can be found in the online repository \citep{TammeEmunds.2023}.

Given the probabilities $p:S \rightarrow [0,1] $ of all states $s \in S$, the expected length of the queue $E_{LW,i}$ of a route $r_i$ can be calculated by summing the probabilities of all states containing elements in the queue
\begin{equation}
\label{equation_queueLength_routes_i}
    E_{LW, i} = \sum_{\Psi_i(s) > 0} p(s) \cdot \Psi(s),
\end{equation}
utilizing the function $\Psi_i: S \rightarrow \mathbb{N}^{0}$, giving the number of elements in the queue of the route $r_i$ for a state $s \in S$. 

Notice that the calculation of the expected length of the queue $E_{LW,i}$ on an arbitrary route $r_i$ in (\ref{equation_queueLength_routes_i}) relies on the use of Markov Chains, solely capable of modelling queueing system of type $M/M/s/\infty$.

In order to analyse systems with general independent ($GI$) arrival or service processes, factors, utilizing the variation coefficient of the described process, can be introduced. According to \citet{KlausFischer.1990}, the expected length of a queue in a Queueing system of type $GI/GI/s/\infty$ can be analysed by using a $M/M/s/\infty$ system and modifying the results 
\begin{equation}
\label{approx_ELW}
    E_{LW, r_i}(M/M/s/\infty) \cdot \frac{1}{\gamma} \approx E_{LW, r_i}(GI/GI/s/\infty)
\end{equation}
accordingly, using
\begin{equation}
\label{approx_ELW_gamma}
    \gamma = \frac{2}{c \cdot v_B^2 + v_A^2}
\end{equation}
and
\begin{equation}
\label{approx_ELW_c}
    c = \left(\frac{\rho}{s}\right)^{1-v_A^2} \cdot (1+v_A^2) -v_A^2.
\end{equation}
Here, $v_A$ and $v_B$ correspond to the coefficients of variation for the arrival and the service process respectively.

Following the equations (\ref{approx_ELW} - \ref{approx_ELW_c}), a greater coefficient of variation in either the arrival or the service process yields a larger expected length of the queue.
In addition to the introduced coefficients of variation, factor $\gamma$ depends on the occupancy rate $\rho$ (see identity (\ref{rho_def})) and the number of parallel service channels $s$. Here, the length of the queue $E_{LW,r_i}$ corresponds to arrivals on one route $r_i$ only, $s=1$ can hence be fixed.

This modification can be implemented in the design of threshold values that have been developed for the capacity analysis of railway infrastructure.

Regarding the modelled railway infrastructure (see Section \ref{section_modelling}), $E_{LW, i}$ has to be calculated for every route $r_i$. Using the threshold values introduced in the next section, every $E_{LW, i}$ can be verified regarding its sufficiency for acceptable operating quality on the infrastructure.

\subsection{Threshold values}
\label{sub_sec:treshold_values}
Different threshold values for theoretical and operational capacity have been discussed in the literature. In the UIC Code 406 \citep{UIC.2013} maximum occupancy rates have been introduced to limit capacity utilization. \citet{GerhartPotthoff.1970} introduces limits for the loss probabilities in railway stations, which are still in practical use for the dimensioning of track numbers in a railway station, i.e. at the German infrastructure manager \cite{DBNetzAG.2009}. 

In this work, the threshold value $L_{W, limit}^{\ast}$, introduced in \citet{Schwanhauer.1982} and likewise still in practical use \citep{DBNetzAG.2009}, is utilized. It corresponds to the maximum number of trains that are to wait in the analysed line section at any given time for a sufficient performance of the infrastructure. 
With the ratio of passenger trains in all considered trains $p_{pt}$, the threshold value
\begin{equation}
\label{wsl_limit_sh}
  L_{W, limit}^{\ast}  = 0.479 \cdot \mathrm{exp}(-1.3 \cdot p_{pt})
\end{equation}
can be specified. 
A threshold value of the approximating queuing system 
\begin{equation}
    L_{W, limit} \approx  \gamma \cdot L_{W, limit}^{\ast} = \gamma \cdot 0.479 \cdot \mathrm{exp}(-1.3 \cdot p_{pt})
\end{equation}
can be obtained by utilizing (\ref{approx_ELW}) and (\ref{wsl_limit_sh}).

Hence, the performance of railway infrastructure is judged based on the arrival and service processes, including their rates and variation, as well as on the operating program during the analysed time horizon.

    

\section{Model Performance}
\label{sec:validation}
Using the formulation of Section \ref{section_modelling}, a Continuous-Time Markov Chain for a railway junction can be obtained. In this work, a maximum of $m=5$ waiting slots per route has been utilized, as a  trade-off between tractability (dependent on the model-size) and accuracy of the model (see Section \ref{sec_example}). By calculating the estimated length of a queue and comparing it to the obtained threshold values (see Section \ref{sub_sec:treshold_values}), the capacity of modelled railway junctions may be assessed. To ensure the quality of the obtained model, the validity of the generated queue-length estimations has to be verified. 

Aiming to survey solely the performance of the solution process using the introduced Continuous-Time Markov Chain, $M/M/s/\infty$ systems have been considered.
For this, simulations of a railway junction with multiple incoming railway lines have been built and run on sample data.

\subsection{Simulation Architecture}

The Simulations have been implemented in Python 3.10.9 \citep{van1995python, PythonSoftwareFoundation.2022}, utilizing SimPy \citep{TeamSimPyRevision.2023}. A model of the junction in Figure \ref{fig:Junctions} has been built, including four different routes with route-specific arrival and service processes.

To estimate inter-arrival times for every route, a pseudo-random number generator yields the next inter-arrival time within a specified exponential distribution with an expected value $ET_{A,r}$ equal to the reciprocal value of the mean arrival rate $\lambda_r = \frac{1}{ET_{A,r}}$ of the modelled route. Trains are stored in a first-in-first-out queue for the route and serviced according to the service time acquired by a second pseudo-random number generator, utilizing another exponential distribution with an expected value $ET_{S}$ equal to the mean service rate $\mu = \frac{1}{ET_{S}}$ of the system.
During the service of a route $r$ it is ensured, that no conflicting route $r^{\prime} \in R$ is able to start service by utilizing shared resources for every pair of routes $(r, r^{\prime}) \in R \times R$.

An implementation can be found in the online repository \citep{TammeEmunds.2023}. To assess the performance of the simulated process, a snapshot of every route's queue-length is taken in every simulated minute. Hence, 
the mean length of a queue can be obtained easily for every simulation run.

\subsection{Validation}

Both, implementations of the simulation and the queueing-length estimations with the formulated Continuous-Time Markov Chain, have been run on a single core of an Intel Xeon Platinum 8160 Processor (2.1 GHz), utilizing a maximum of 3900 MB working memory. 

Two different simulations setups have been considered:
\begin{enumerate}
    \item A simulation with no limit to the number of trains being able to wait in the queues of their requested routes
    \item A simulation with a maximum of 5 trains per route being able to wait at the same time for the service on their requested routes
\end{enumerate}

In the second simulation and in the analytical setting, an arriving train is rejected, i.e. not inserted into the respective queue, if its arrival time lies within a time frame where 5 trains are already waiting for the release of the requested route. Here, all routes are capable of having 5 trains waiting for service, hence a total maximum of 20 trains may wait at the same time.

All three solution approaches, both simulation setups and the analytical method, have been set to compute the estimated length of the queue $E_{LW,r}$ for the route $r_3$ (see Table \ref{tab:routes}), conflicting with routes in direction of A and C (see Figure \ref{fig:Junctions}). Route $r_3$ has been selected as it is one of the routes with the most conflicts (with $r_2$ and $r_4$) and it would directly benefit from an overpass construction. 

Since those computations have been done to compare the results and running times of the approaches, only a small set of 10 different service rates, between 0.1 and 1.0 trains per minute, has been considered. An arrival rate of 0.1 trains per minute has been set for every route in all computed instances.

Each simulation has been run 100 times for every considered service time, simulating a total of 22 hours for every run. From those 22 hours, the first and last have not been considered, resulting in an evaluated time of 20 hours per simulation and 2000 hours in total for every service time investigated.

\begin{figure}[tp]
\centering
\includegraphics[width=.9\linewidth]{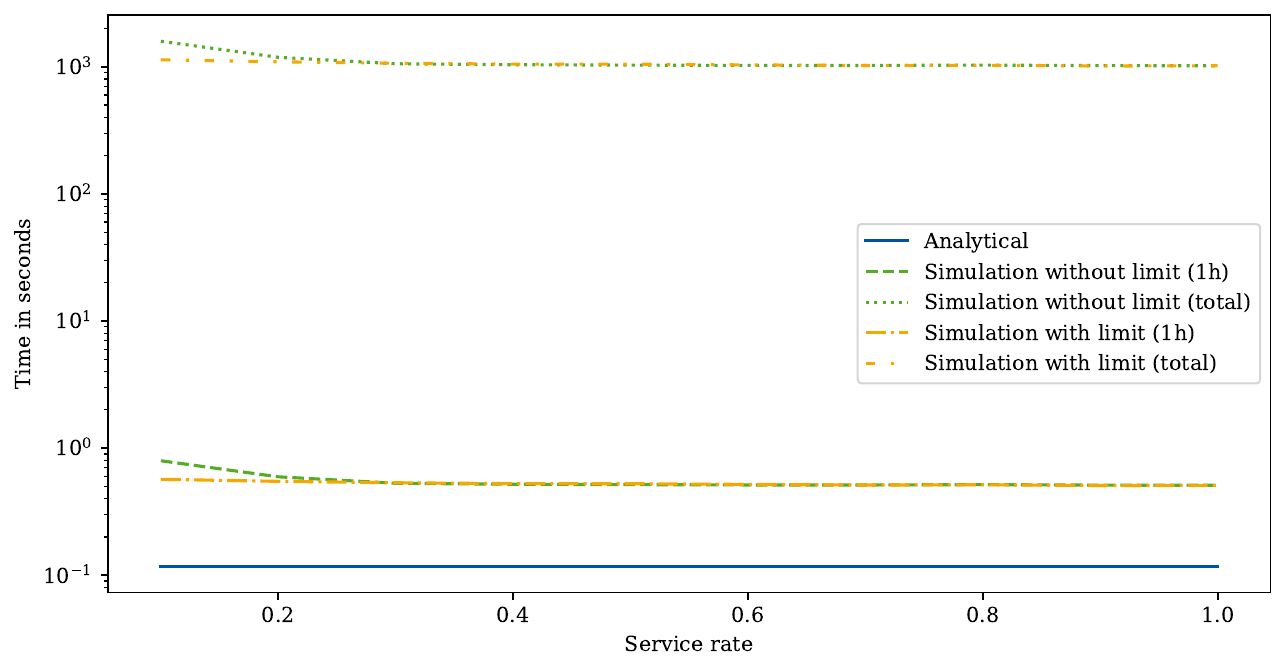}
\captionsetup{justification=centering}
\caption{Computation times needed for the analytical compared to the simulation approach}
\label{fig:sim_times}
\end{figure}

For both simulation setups the mean computing time per hour and the mean total run time per 22 hour simulation have been evaluated. Those can be found in Figure \ref{fig:sim_times}, which additionally includes the running time of the analytical solution for reference.

The computing time results in Figure \ref{fig:sim_times} clearly indicate that the analytical approach is faster, even if compared to the run of a single simulation hour. Noting the logarithmic scale, for small service rates the analytical approach is faster by a factor of 5 to 10, depending on the considered service rate. Since simulations have to be conducted multiple times in order to receive sufficient results, a comparison with the total simulation times might be more realistic, yielding factors of $10^3$ up to $10^4$.

In general, the computing times of the simulation approach are significantly increasing with a decreasing service rate. This is probably due to a higher amount of trains in the network at any given time, as the mean service time increases with a decreasing service rate, leading to more trains waiting for the release of their requested route. Contrary, simulation runs with a setup without limit required almost the same computing time compared to simulations subject to a maximum of 5 trains per queue. 

\begin{figure}[th]
\centering
\begin{subfigure}[t]{.5\textwidth}
  \centering
  \includegraphics[width=.9\linewidth]{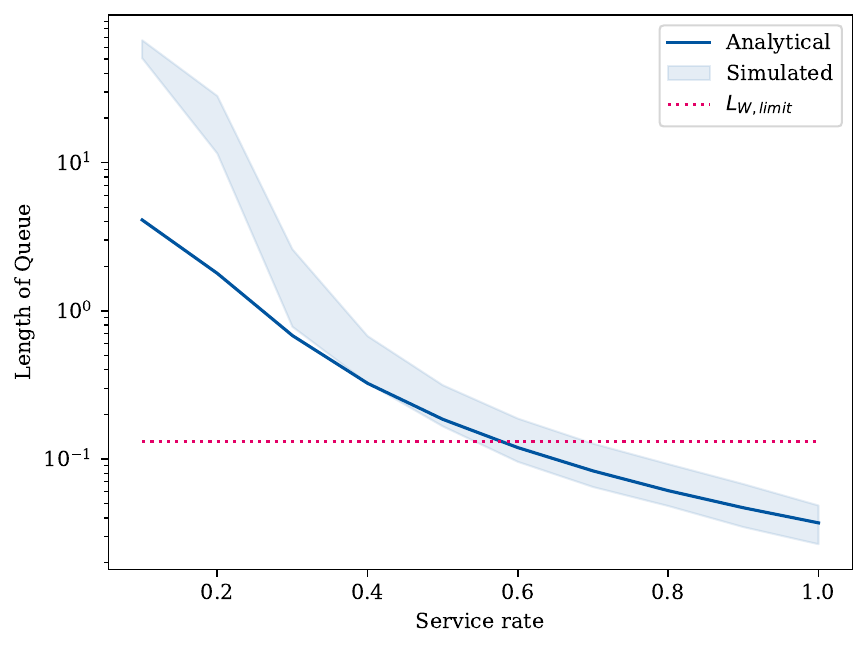}
  \captionsetup{justification=centering}
  \caption{Simulation with no limit on the length of a queue}
  \label{fig:sim_performance_ol_plot}
\end{subfigure}%
\begin{subfigure}[t]{.5\textwidth}
  \centering
  \captionsetup{justification=centering}
  \includegraphics[width=.9\linewidth]{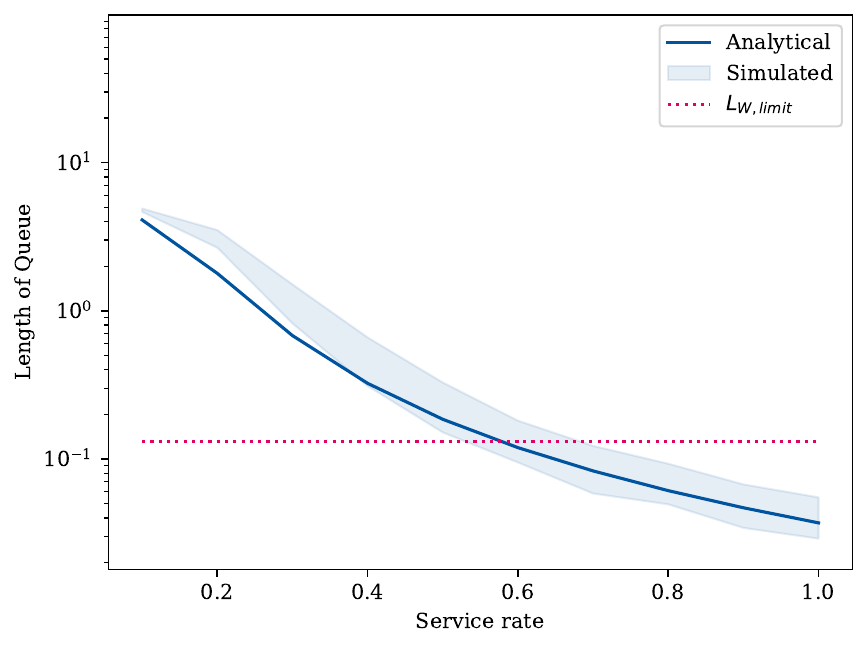}
  \caption{Simulation with a limit of $m = 5$ on the length of a queue}
  \label{fig:sim_performance_l_plot}
\end{subfigure}
\caption{Accuracy of the analytical approach in comparison to the confidence interval of conducted simulations}
\label{fig:sim_performance}
\end{figure}

Additional observations can be made by evaluating the accuracy of the introduced analytical approach. This can be done by comparing the obtained results of the simulation methods with the results of the queueing-based analytical approach. In Figure \ref{fig:sim_performance} the obtained results of the $E_{LW, r_3}$ computations are depicted by including the exact analytical results and the standard deviation area of the simulation results, i.e. the area $\left[\overline{E}_{LW} - \sigma, \overline{E}_{LW, r_3} + \sigma \right]$, surrounding the mean $\overline{E}_{LW, r_3}$ by the standard deviation $\sigma$. 

Noting the logarithmic scale in Figure \ref{fig:sim_performance}, the results show significant differences between the simulation setups. For the setup without a limit on the number of trains in a queue (Figure \ref{fig:sim_performance_ol_plot}), serious discrepancies between the simulation and the analytical can be found for service rates of less than 0.3. Results of the analytical solution seem to be more similar to the results of the simulation setup with a limit of $m = 5$ on the length of a queue (Figure \ref{fig:sim_performance_l_plot}), matching the setting for the queueing-based analytical approach.
Hence, an assumption, that for railway junctions, utilizing the fixed predefinitions and an averaged service rate of less than 0.3, the number of trains in a queue will on average exceed the limit of 5, can be made. Consequently, arriving trains are likely to have to wait, resulting in a very poor service quality.

Taking the limit for $M/M/s/\infty$ queueing systems (see Section \ref{sub_sec:treshold_values}), in Figure \ref{fig:sim_performance} with '$L_{W, limit}$' annotated, into account, the accuracy of the introduced queueing-based analytical approach seems to be sufficient for the use in practical applications.




\section{Computational Study}
\label{sec:Computational_Study}

The introduced method for the calculation of queuing lengths can be used to guide infrastructure managers when dimensioning railway junctions. This work particularly focuses on choosing the right infrastructure layout for a given operating program, i.e. problem statement (II) in Section \ref{subsection_problem_formulation}.

\subsection{Setup}

\begin{figure}[ht]
    \centering
    \includegraphics[width=.9\linewidth]{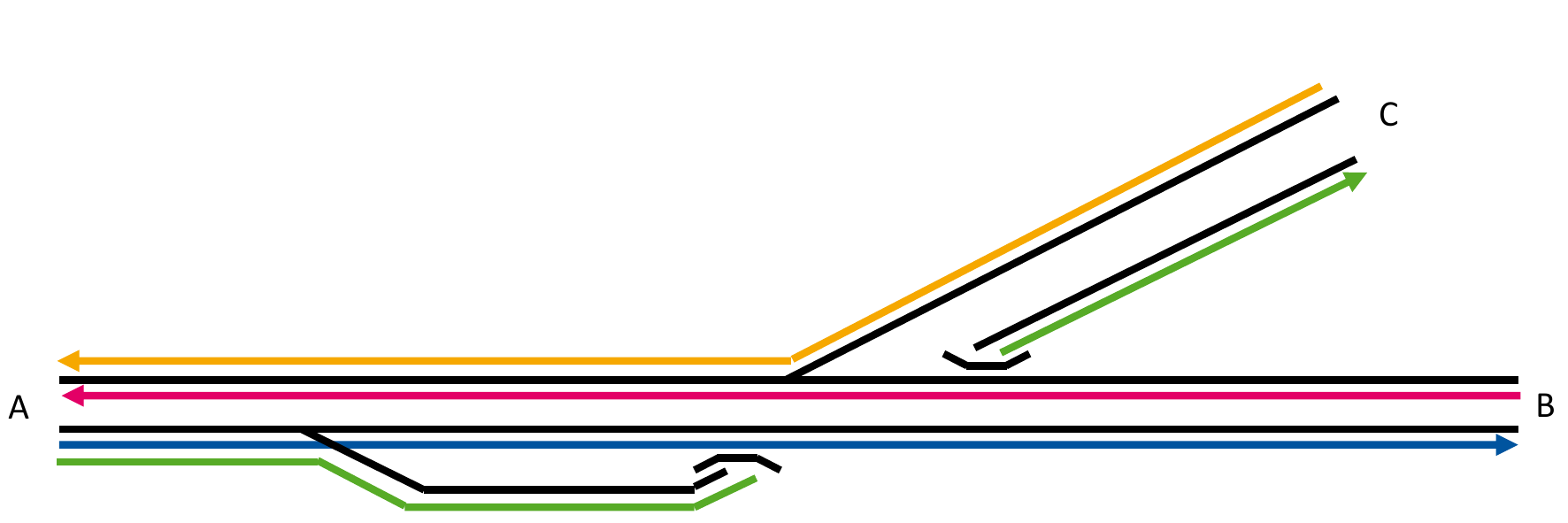}
    \caption{Track-layout for a double-track junction with an overpass}
    \label{fig:Junction_overpass}
\end{figure}

In detail, a case study, deciding whether or not an overpass should be built for a junction with a given operating program has been conducted. An overpass is a way to reduce the number of route conflicts in a railway junction, an example for the track layout of the junction in Figure \ref{fig:Junctions} is given in Figure \ref{fig:Junction_overpass}.

\begin{figure}[ht]
    \centering
    \resizebox{1.0\textwidth}{!}{
    \includegraphics{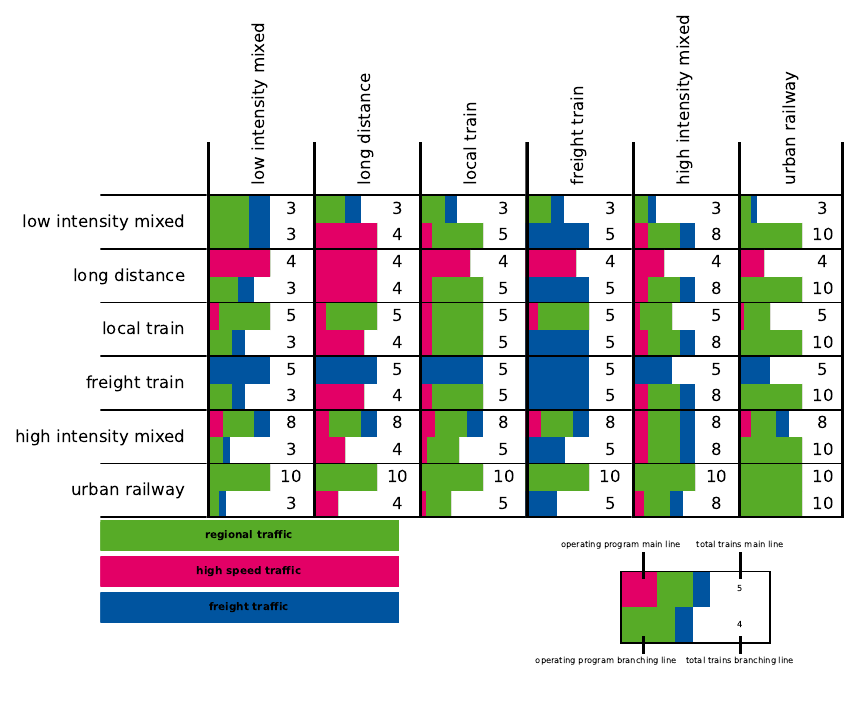}
    }
    \caption{Operating program combinations considered 
    }
    \label{fig:operating_programme}
\end{figure}

To show the applicability of the described analytical approach, 23 different operating programs have been considered. Thus, all combinations of 23 different operating programs for both to the railway junction adjacent railway lines have been built and analysed for their peak traffic hour, resulting in a total number of $23^2 = 529$ different examples. 
A detailed description of all considered operating program combinations can be found in \ref{app:comp_study}, we restrict ourselves to six exemplary railway lines here.

The operating programs have been selected according to different types of the adjacent railway lines, we distinguish between \textit{mixed traffic lines}, \textit{local train lines}, \textit{long-distance train lines}, \textit{freight train lines}, and \textit{urban railway lines}. Additionally, two different loads have been considered for the mixed traffic line. Table \ref{tab:op_programs} lists the selected operating programs.

\begin{table}[h]
    \centering
    \caption{Operating programs considered}
    \begin{tabular}{c|c|c|c}
         \makecell{operating \\ program} &  \makecell{\# regional \\ trains}   & \makecell{\# high speed \\ trains} & \makecell{\# freight \\ trains}\\ \hline
         low intensity mixed & 2 & 0 & 1 \\ 
         long distance  & 0 & 4 & 0 \\ 
         local train & 4 & 1 & 0 \\ 
         freight train  & 0 & 0 & 5 \\ 
         high intensity mixed  & 4 & 2 & 2 \\ 
         urban railway & 10 & 0 & 0 \\ \hline
    \end{tabular}
    \label{tab:op_programs}
\end{table}


In Figure \ref{fig:operating_programme} the considered combinations of those operating programs have been listed. In every entry, the top bar corresponds to the operating program on the \textit{main line}, i.e. the routes from A to B (\textcolor{rwth}{$r_1$}) and from B to A (\textcolor{magenta}{$r_3$}), while the bottom bar corresponds to the operating program on the \textit{branching line}, i.e. the routes from (\textcolor{orange}{$r_4$}) and to (\textcolor{grun}{$r_2$}) C.

The total number of trains on a route $n_r$ is utilized to determine the arrival rate $\lambda_r = \frac{1}{n_r}$ for this route. Next, Continuous-Time Markov Chains (see Section \ref{section_modelling}) can be formulated for any possible service rate $\mu \in (0,1]$, using the defined arrival rates $\lambda_r$ and the maximum number of waiting positions per queue $m=5$, while also taking conflicting routes into consideration. To ensure a sufficient precision while also maintaining efficient computability, all service rates $\mu \in [0.01,1]$ with a step size of 0.01 have been taken into consideration in this work.

Hence, 100 Models have been analysed for both junction infrastructure layouts with/without an overpass for every considered combination of operating programs. The solving process has been automated using Python 3.10.9 \citep{van1995python, PythonSoftwareFoundation.2022} and state-of-the-art model-checking Software. For this work, the model-checker \textit{Storm} \citep{Hensel.2022} has been utilized. With its Python-Interface \textit{Stormpy} \citep{SebastianJunges.2023} it is easy to use while also accomplishing competitive results in qualitative benchmarks \citep{Budde.2021}. Used Models have been formulated according to Section \ref{section_modelling}, expressed in the PRISM modelling language \citep{DaveParkerGethinNormanMartaKwiatkowska.2000, KNP11}. Detailed information regarding the solving process and the used model files can be found in the online repository \citep{TammeEmunds.2023}.

Aiming to gain insights about the performance benefit of the infrastructure layout including the overpass (see Figure \ref{fig:Junction_overpass}), the expected length of the queue at route \textcolor{magenta}{$r_3$}, from B to A, has been analysed. It has been chosen as it has two conflict points with other routes, one with the other route to A, \textcolor{orange}{$r_4$} from C, the other with the route \textcolor{grun}{$r_2$} from A to C, which is only conflicting in the junction layout without an overpass structure.

\begin{figure}[ht]
    \centering
    \includegraphics[width=.7\linewidth]{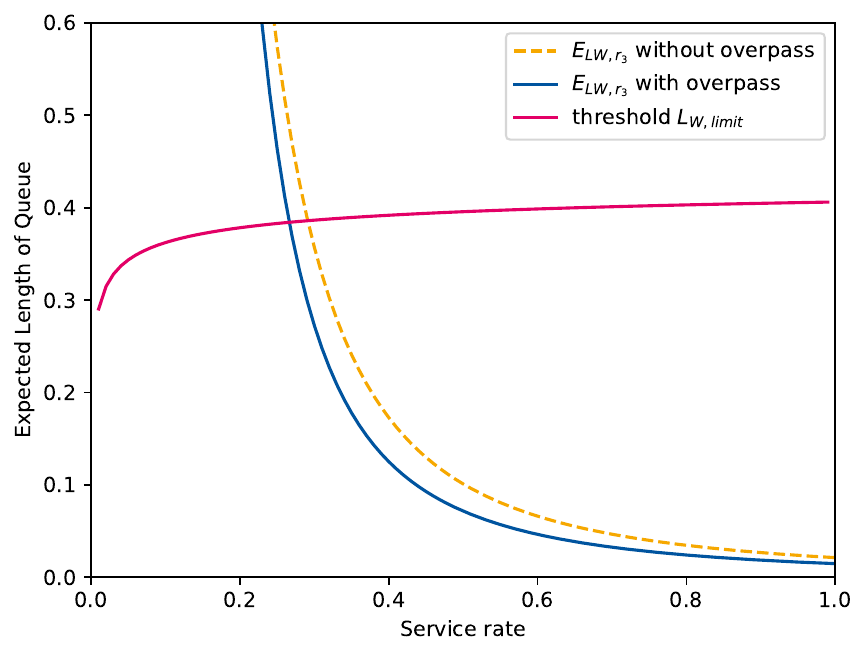}
    \caption{Estimated lengths of the queue on route \textcolor{magenta}{$r_3$} for the combination of a local train main line and a long-distance train branching line}
    \label{fig:p3_p1}
\end{figure}

Resulting from the solving process is a grid of expected lengths of the queue for both infrastructure layouts for every considered service rate $\mu \in [0.01,1]$.
In Figure \ref{fig:p3_p1} the expected length of the queue at route \textcolor{magenta}{$r_3$} from B to A for both railway junction layouts of a local train main line and a long-distance train branching line is recorded.
Both operating programs are only considering passenger transport, hence a ratio of passenger trains in all considered trains of $p_{pt} = 1$ can be used for the calculation of the threshold value $L_{W, limit}$.

The combination is assumed to have to deal with an hourly load of 5 trains on the main line, from A to B and back, as well as 5 additional trains on the branching line, from A to C and back.
This resumes in an arrival rate of $\lambda_r = 0.083$ for the routes $r \in \{r_1, r_3\}$ and of $\lambda_{r^{\prime}} = 0.067$ for the routes $r^{\prime} \in \{r_2, r_4\}$.

Furthermore, the threshold $L_{W, limit}$ is depicted, modified according to Section \ref{sub_sec:treshold_values}, depending on the occupancy rate of $\rho = \frac{\lambda_{r_3}}{\mu}$ for route \textcolor{magenta}{$r_3$}. In this work, the coefficients of variation are assumed to be $v_A = 0.8$ for the arrival process and $v_S = 0.3$ for the service process, according with standard values in literature (see \citet{Wendler.1999}).

Utilizing the grid of resulting queue-length estimations as well as the calculated threshold, a minimum mean service rate $\mu_{\text{min}}$, needed for sufficient infrastructure quality, may be obtained. Hence, the maximum mean service time $b_{\text{max}} = \frac{1}{\mu_{\text{min}}}$ for the given operation program on the main and branching lines can be derived. This maximum mean service time can be used to investigate the needed infrastructure by comparing it to the actual achieved service times on the analysed junction, which are subject to train and control system specific parameters.

\subsection{Results}
\label{sec:results}

The introduced derivation of a maximum mean service time has been applied to all 529 considered operating program combinations for the infrastructure settings with and without an overpass. 

\begin{figure}[htp]
    \centering
    \resizebox{0.85 \textwidth}{!}{
    \includegraphics{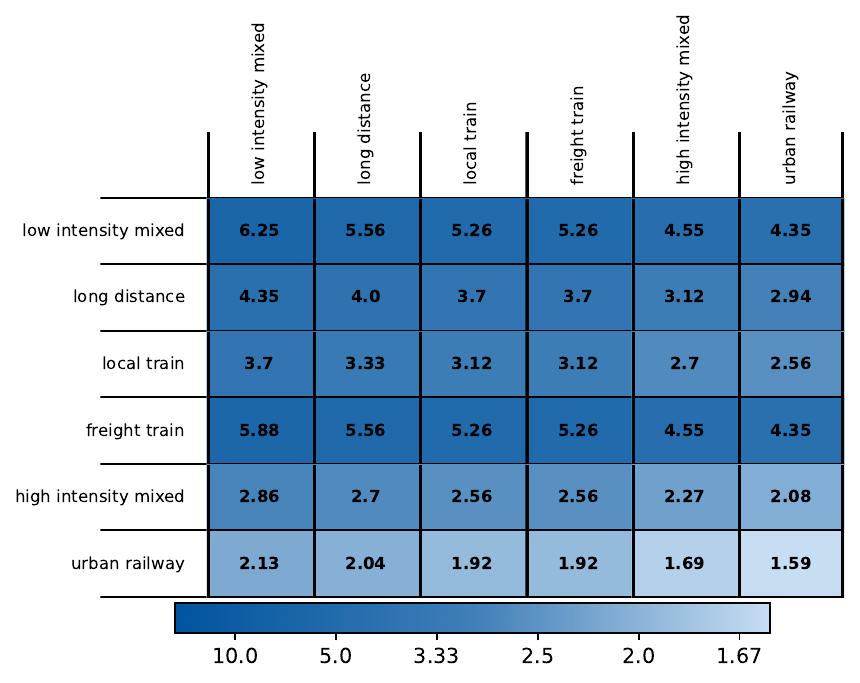}
    }
    \caption{Resulting maximum mean service times for the considered operating program combinations without an overpass}
    \label{fig:results_heatmap_without}
\end{figure}

\begin{figure}[htp]
    \centering
    \resizebox{0.85 \textwidth}{!}{
    \includegraphics{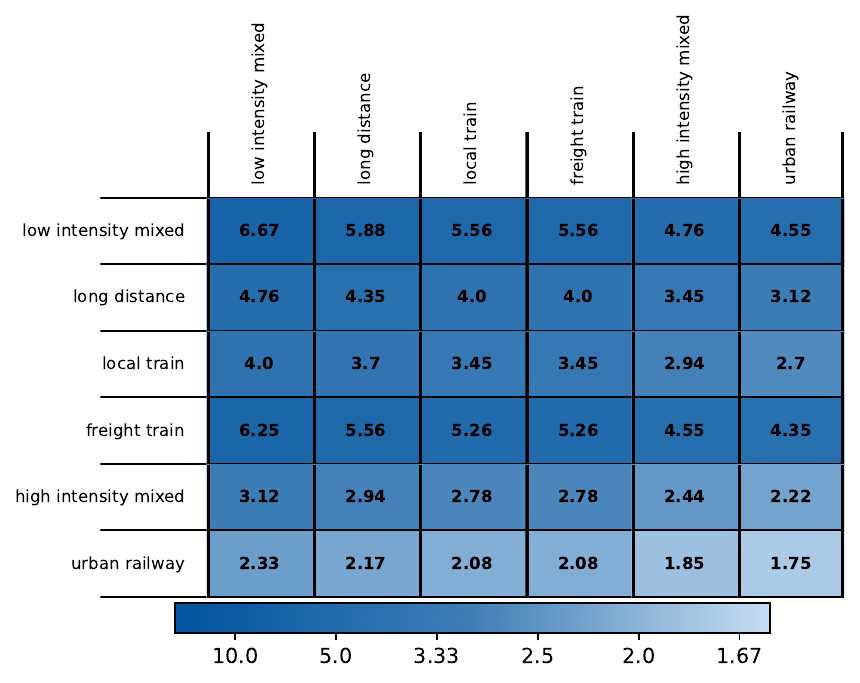}
    }
    \caption{Resulting maximum mean service times for the considered operating program combinations with an overpass}
    \label{fig:results_heatmap_with}
\end{figure}

In Figures \ref{fig:results_heatmap_without} and \ref{fig:results_heatmap_with} the results for the selected 36 combinations are shown in heat-maps for both considered railway junction layouts. While both infrastructure settings share the same global distributions of service time requirements, results indicate that under the same load, as fixed by the operational program of the main and branching line, a railway junction including an overpass structure allows for a higher mean service time, while achieving the same operational quality as a railway junction without the overpass structure.

\begin{figure}[th]
\centering
\resizebox{1.\textwidth}{!}{
\begin{subfigure}[b]{.7\textwidth}
  \centering
  \includegraphics[width=1.\linewidth]{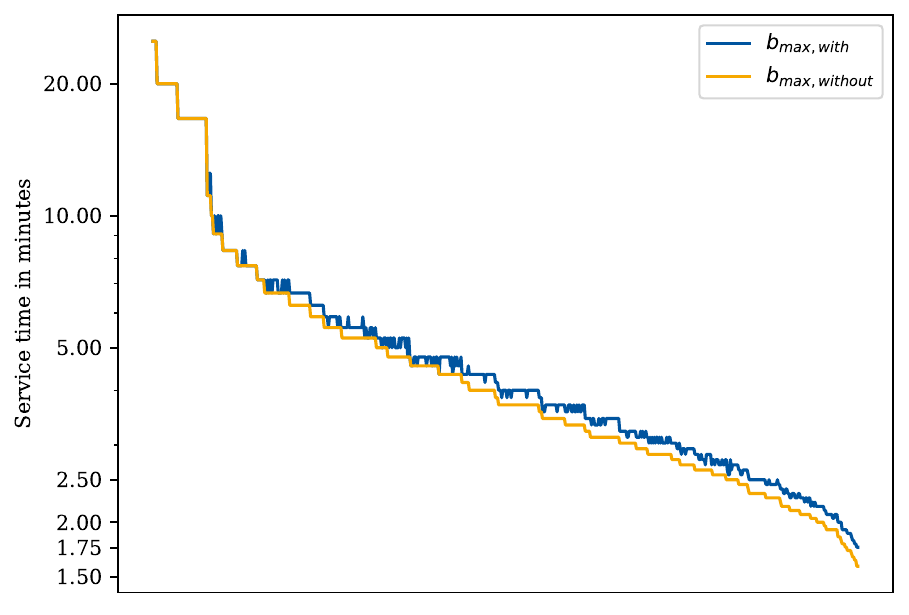}
  \captionsetup{justification=centering}
  \caption{Resulting maximum mean service times for both infrastructure settings}
  \label{fig:result_distribution_both}
\end{subfigure}%
\begin{subfigure}[b]{.3\textwidth}
  \centering
  \captionsetup{justification=centering}
  \includegraphics[width=.84\linewidth]{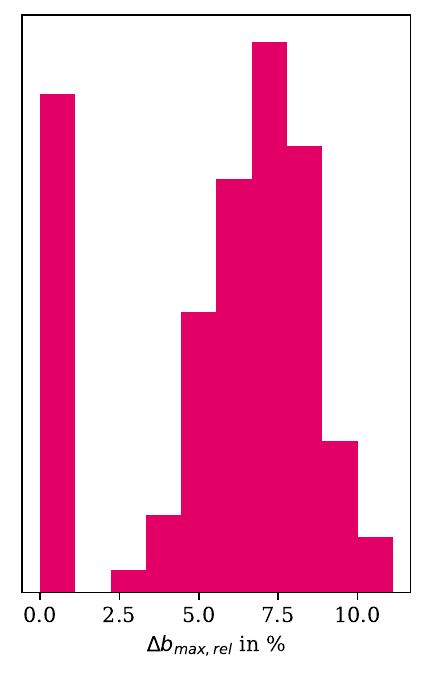}
  \caption{Relative difference in maximum mean service times}
  \label{fig:result_distribution_hist}
\end{subfigure}
}
\caption{Comparison of the obtained maximum mean service times}
\label{fig:result_distribution}
\end{figure}

Further investigating the difference between the two infrastructure settings, Figure \ref{fig:result_distribution} includes the maximum mean service times (Figure \ref{fig:result_distribution_both}) for all 529 computed combinations of main and branching lines (see also \ref{app:comp_study}) and a histogram, representing the distribution of relative differences in the calculated maximum mean service times (Figure \ref{fig:result_distribution_hist}).

Taking the logarithmic scale in Figure \ref{fig:result_distribution_both} into account, a wide range of calculated maximum mean service times required for sufficient operational quality can be recognized. While some operating programs, i.e. examples with a low number of total trains in the considered peak hours, are granting sufficient operational quality even for mean service times as high as 10 to 20 minutes, the majority of considered examples require a mean service time of under 5 minutes, with very densely operated main and branching lines demanding maximum mean service times of even under 2 minutes.

By including an overpass in the layout of a planned railway junction, the required maximum mean service times can be decreased by a significant margin. For the majority of considered examples, the relative difference in the maximum mean service time lies between 5\% and 10 \%, but the achieved reduction of maximum mean service time is diverse. Some examples show virtually no difference between the considered infrastructure layouts, others report maximum mean service time decreases by up to 10 \%.

Crucially, service times are dependent on various factors, some of which are not under the influence of infrastructure managers. Hence, substantial limitations to the lower bound of service times are relevant, affected by physical properties such as length or acceleration and braking performance of the rolling stock.

Concluding the computational experiment, differences between the considered infrastructure layouts have been analysed with the introduced method. For this, the calculation of many different operating program and service rate configurations has been conducted, resulting in substantiated estimations of infrastructure quality requirements. The achieved results indicate that railway junctions with a high total number of trains per hour can benefit from overpass structures to resolve some conflicts on requested routes.

\section{Discussion and Outlook}
\label{sec:Discussion}

This work introduced a novel method for analysing the timetable capacity of railway junctions based on queueing theory. It is applicable for solving both formulated problem statements (Section \ref{subsection_problem_formulation}), timetable capacity determination of a given railway junction infrastructure (I) and dimension of junction infrastructure for a fixed operating program (II). By modelling railway junction routes as parts of a queueing system, while respecting their parallel service possibilities and taking resource conflicts into consideration, timetable independent analyses of examined infrastructure are enabled.

Utilizing classical queueing theory concepts, well established for railway line performance analysis (see Section \ref{sec:queueing_system}), timetable capacity is determined by comparing queue-length estimations through a Continuous-Time Markov Chain representing the considered railway junction (see Section \ref{section_modelling}) with threshold values (see Section \ref{sub_sec:treshold_values}) depending on parameters of considered service and arrival process distributions along with operating program specifics. In this work, estimations of queue lengths have been carried out by model-checking software, enabling a fast and reliable computation for complex infrastructure dependencies.

The performance of the introduced approach has been studied by exemplary comparing the computation times and results of the analytical solution with simulations (see Section \ref{sec:validation}). Those simulations indicate that the accuracy of the introduced method does fulfill the requirements for sufficient analysis when utilizing threshold values, while also being significantly faster than the applied simulations.

Implementing the novel analytical model for operating programs in computational experiments (see \ref{sec:Computational_Study}), the selection of sufficient junction infrastructure has been tested. For this, 529 different operating program combinations have been considered, leading to substantive estimations regarding the effect of overpass structure on the timetable capacity of railway junctions. Hence, the introduced method has been proven to be applicable to the solution of conceptional issues on abstract and implementing scales.

Even though the concept is already applicable, additional research could still yield substantial benefit. As such, the modelling of general independent stochastic processes with Markov processes and approximation factors might be improvable, i.e. by including phase-type distributions.
Similarly, utilized thresholds for the expected length of a queue have been introduced for the use in a context of railway lines, updating these to innovative measures could improve the real world applicability. 
Additionally, the introduced concept could be extended to implementations for railway stations and eventually railway networks. By including delay distributions and propagation in the model, similar measures for operational capacity could be enabled. 

Utilizing the developed timetable independent method, infrastructure managers are however able to identify bottlenecks in early stages of the planning process.
With the achieved computing time benefits when compared to simulation approaches, they might be setup to give junction designers substantiated indicators for required infrastructure.
Hence, the introduced analytical timetable independent approach might be proven to form a valuable addition to the capacity method landscape.








\section*{CRediT authorship contribution statement}
\textbf{Tamme Emunds:} Conceptualization, Methodology, Software, Formal Analysis, Writing – original draft. \textbf{Nils Nießen:} Conceptualization, Supervision, Writing - review \& editing.

\section*{Declaration of competing interest}
The authors declare that they have no known competing financial interests of personal relationships that could have appeared to influence the work reported in this paper.

\section*{Acknowledgements}
The authors thank Mr. Alexander Bork for his invaluable insights regarding model-checking techniques in general and regarding the software Storm in specific. Furthermore, the authors thank Mr. Tobias Müller and Dr. Andreas Pfeifer for their supervision and guidance regarding practical implementations as well as providing application relevant case examples. Additionally, the authors thank DB Netz AG for the opportunity of applying the described theory in a practical project.

This work is funded by the Deutsche Forschungsgemeinschaft (DFG, German Research Foundation) – 2236/2. Computational Experiments were performed with computing resources granted by RWTH Aachen University under project rwth1413.

\bibliographystyle{elsarticle-harv}
\bibliography{references}

\newpage
\appendix

\section{Literature Summary}
\label{app:lit_ges}

\begin{table}[h]
\centering
\caption{Literature considering railway performance estimations}
\label{table:lit_all}
\resizebox{0.84 \textwidth}{!}{
\rotatebox{90}{
\begin{tabular}{c|c|c|c|c|c|c}
reference                    & capacity type & timetable dependence & methodology               & infrastructure type & infrastructure decomposition  & solution technique       \\ \hline \hline
\citet{UIC.2004}             & utilization   & dependent            & compression               & line                & block-section                 & closed-form formula      \\
\citet{UIC.2013}             & utilization   & dependent            & compression               & line and junction   & block-section and route-based & closed-form formula      \\
\citet{Abril.2008}           & utilization   & dependent            & compression               & line                & block-section                 & closed-form formula      \\
\citet{Landex.2009}          & utilization   & dependent            & compression               & line                & block-section                 & closed-form formula      \\
\citet{Landex.2013}          & utilization   & dependent            & compression               & junction            & block-section                 & closed-form formula      \\
\citet{Goverde.2013}         & utilization   & dependent            & compression               & line                & train-path                    & closed-form formula      \\
\citet{Jensen.2017}          & utilization   & independent          & compression               & single-line network & train-path                    & combinatorial algorithms \\
\citet{Jensen.2020}          & utilization   & independent          & compression               & single-line network & train-path                    & combinatorial algorithms \\
\citet{Weik.2020}            & utilization   & independent          & compression               & station             & train-path                    & closed-form formula      \\ \hline
\citet{Goverde.2007}         & utilization   & dependent            & max-plus-algebra          & network             & timed-events                  & matrix-calculations      \\
\citet{Besinovic.2018}       & utilization   & dependent            & max-plus-algebra          & network             & timed-events                  & matrix-calculations      \\
\citet{Kort.2003}            & operational   & independent          & max-plus-algebra          & single-line network & timed-events                  & matrix-calculations      \\ \hline
\citet{Harrod.2009}          & theoretical   & timetable-saturation & optimisation              & line                & train-path                    & MIP                      \\
\citet{Yaghini.2014}         & theoretical   & independent          & optimisation              & line                & timed-events                  & MIP                      \\
\citet{Zwaneveld.1996}       & theoretical   & independent          & optimisation              & station             & route-based                   & MIP                      \\
\citet{Zwaneveld.2001}       & theoretical   & independent          & optimisation              & station             & route-based                   & MIP                      \\
\citet{Delorme.2001}         & theoretical   & independent          & optimisation              & junction            & block-section                 & combinatorial algorithms \\
\citet{Burdett.2006}         & theoretical   & timetable-saturation & optimisation              & network             & train-path                    & MIP                      \\
\citet{Burdett.2015}         & theoretical   & independent          & optimisation              & network             & train-path                    & combinatorial algorithms \\
(\citet{Cacchiani.2012})     & (theoretical) & independent          & optimisation              & -                   & -                             & mostly MIP               \\
\citet{Leutwiler.2022}       & (theoretical) & independent          & optimisation              & network             & timed-events                  & disjunctive programming  \\
\citet{Zhang.2016}           & utilization   & independent          & optimisation              & line                & timed-events                  & combinatorial algorithms \\
\citet{Liao.2021}            & theoretical   & timetable-saturation & optimisation              & line                & timed-events                  & MIP                      \\
\citet{Mussone.2013}         & utilization   & independent          & optimisation              & network             & train-path                    & MIP                      \\ \hline
\citet{Graffagnino.2012}     & operational   & dependent            & operational-data-analysis & line                & train-path                    & statistical analysis     \\
\citet{Armstrong.2017}       & utilization   & dependent            & operational-data-analysis & station             & timed-events                  & statistical-analysis     \\
\citet{Weik.2022}            & operational   & dependent            & operational-data-analysis & line                & block-section                 & statistical-analysis     \\
(\citet{Corman.2022})        & operational   & dependent            & operational-data-analysis & line                & train-path                    & statistical-analysis     \\ \hline
\citet{Goverde.2010}         & (operational) & dependent            & max-plus-algebra          & network             & timed-events                  & matrix-calculations      \\
\citet{Buker.2011}           & (operational) & independent          & analytical                & network             & timed-events                  & iterative formula        \\
\citet{Sahin.2017}           & (operational) & dependent            & analytical                & line                & timed-events                  & matrix-calculations      \\
\citet{Zieger.2018}          & (operational) & independent          & simulation                & line                & train-path                    & simulation               \\
\citet{Corman.2018}          & (operational) & dependent            & operational-data-analysis & line                & timed-events                  & bayesian-network         \\
(\citet{Spanninger.2023})    & (operational) & -                    & operational-data-analysis & -                   & -                             & -                        \\  \hline
\citet{GerhartPotthoff.1970} & timetable     & independent          & analytical                & station             & route-based                   & iterative formula        \\
\citet{Schwanhauer.1974}     & operational   & independent          & analytical                & line                & train-path                    & closed-form formula      \\
\citet{Schwanhauer.1978}     & timetable     & independent          & analytical                & junction            & route-based                   & closed-form formula      \\
\citet{Wakob.1984}           & timetable     & independent          & analytical                & line                & train-path                    & closed-form formula      \\
\citet{Wendler.2007}         & timetable     & independent          & analytical                & line                & train-path                    & matrix-calculations      \\
\citet{Niessen.2008}         & operational   & independent          & analytical                & junction            & individual                    & iterative formula        \\
\citet{Niessen.2013}         & timetable     & independent          & analytical                & junction            & individual                    & iterative formula        \\
\citet{Schmitz.2017}         & timetable     & independent          & analytical                & junction            & individual                    & matrix-calculations      \\
\citet{Weik.2020PhD}         & timetable     & independent          & analytical                & line and junction   & route-based                   & matrix-calculations     \\ \hline
Introduced here              & timetable     & independent          & analytical                & junction            & route-based                   & model-checking          \\ \hline
\multicolumn{7}{p{1.4\textheight}}{Remarks: Please note that literature referring to the \textit{(operational)} capacity type is introducing methods for computing delay propagation, which is the basis for calculating operational capacity.
Furthermore, literature referring to the \textit{(theoretical)} capacity type is introducing methods for railway timetabling, which can be used to compute theoretical capacity.
Literature written within bracelets \textit{(reference)} does not introduce new methods but rather gives a comprehensible review of relevant methodology.}
\end{tabular}
}
}
\end{table}

\newpage
\section{Full Computational Study}
\label{app:comp_study}

The computational study explained in section \ref{sec:Computational_Study} has been conducted for 529 operating program combinations in total. Since we refrained from showing their results in full detail in the text, some additional figures are given on the next pages. Please note that the results can additionally be found in the online repository \citep{TammeEmunds.2023}.

Table  \ref{tab:op_selected} lists all selected operating programs in section \ref{sec:Computational_Study} and their respective names in the following figures.

\begin{table}[h]
\centering
\caption{Selected operating programs}
\label{tab:op_selected}
\begin{tabular}{l|l|l|l}
Name of railway line  & Name in appendix       \\ \hline
long-distance         & P1, (0)           \\
local train           & P3, (0)     \\
urban railway         & P3-P5, (+)       \\
freight train         & F1, (0)   \\
low intensity mixed   & P3\_F1, (0)       \\
high intensity mixed  & P3\_F1, (+)  
\end{tabular}
\end{table}

\begin{figure}[htp]
    \centering
    \resizebox{1.\textwidth}{!}{
    \includegraphics[angle=90]{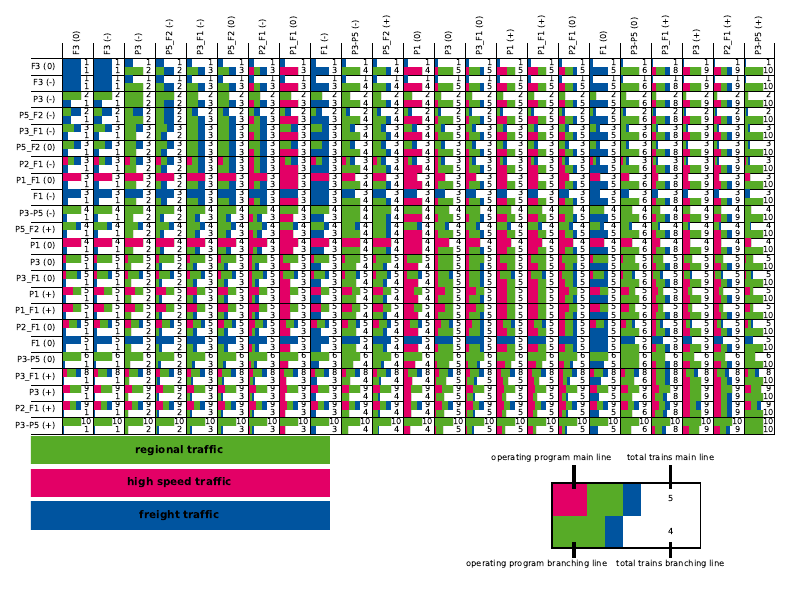}
    }
    \caption{Operating program combinations considered 
    }
    \label{fig:operating_programme_all}
\end{figure}

\begin{figure}[htp]
    \centering
    \resizebox{1.0\textwidth}{!}{\includegraphics[angle=90]{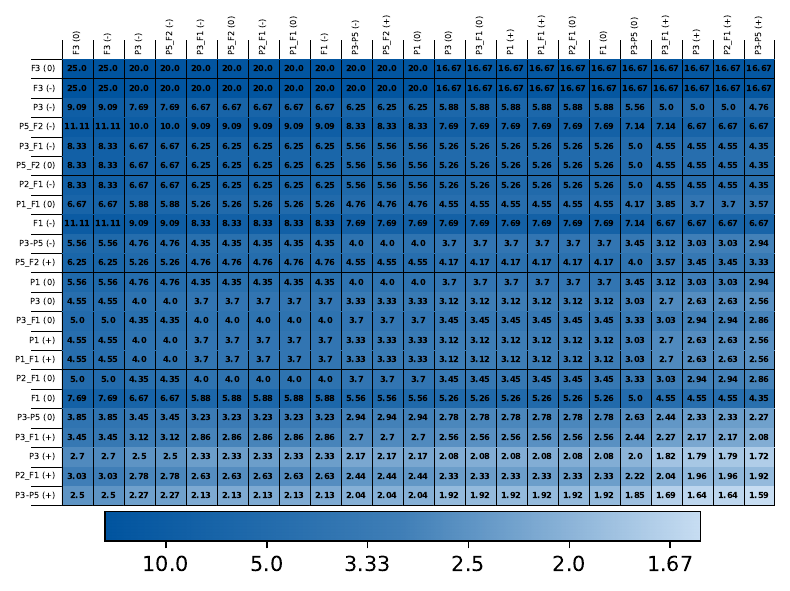}}
    \caption{Resulting maximum mean service times for the considered operating program combinations without an overpass}
    \label{fig:results_heatmap_without_all}
\end{figure}
\begin{figure}[htp]
    \centering
    \resizebox{1.0\textwidth}{!}{\includegraphics[angle=90]{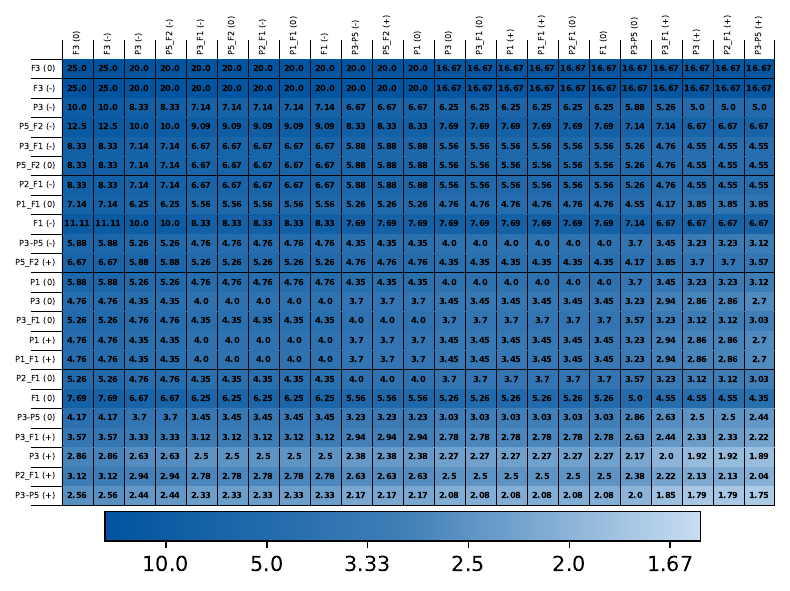}}
    \caption{Resulting maximum mean service times for the considered operating program combinations with an overpass}
    \label{fig:results_heatmap_with_all}
\end{figure}

In Figure \ref{fig:operating_programme_all} all used operating program combinations are depicted. Considered operating programs have been created as samples for demands to railway infrastructure and named in accordance with \citep{EuropeanComission.16May2019}, additionally distinguished between the three categories of exposure \textit{base load} (-), \textit{medium load} (0) and \textit{high load} (+). Table \ref{tab:traffic_codes} lists a selection of traffic code parameters as specified in \cite{EuropeanComission.16May2019}.

\begin{table}[h]
\centering
\caption{Selected Traffic code parameters from \citep{EuropeanComission.16May2019}}
\label{tab:traffic_codes}
\begin{tabular}{c|c|c|c|c}
Traffic Code & Traffic type & \makecell{Line Speed \\ $[\text{km/h}]$ \, } & \makecell{Usable length of  \\ platform $[\text{m}]$}  & \makecell{Train length \\ $[\text{m}]$} \\ \hline
P1 & passenger traffic & 250-350 & 400 & n.a. \\
P2 & passenger traffic & 200-250 & 200-400 & n.a.\\
P3 & passenger traffic & 120-200 & 400-400 & n.a. \\
P5 & passenger traffic & 80-120 & 50-200 & n.a.\\ \hline
F1 & freight traffic & 100-120 & n.a. & 740-1050 \\
F2 & freight traffic & 100-120 & n.a. & 600-1050\\
F3 & freight traffic & 60-100 & n.a. & 500-1050 \\

\end{tabular}
\end{table}

Additionally, the corresponding results are depicted in Figures \ref{fig:results_heatmap_without_all} and \ref{fig:results_heatmap_with_all} for the calculations of maximum mean service times for the two infrastructure settings without and with an overpass.

\end{document}